\documentclass[useAMS,usenatbib]{mn2e}
\usepackage{graphicx}

\newcommand{\AlII}{Al\,\textsc{ii}}
\newcommand{\AlIII}{Al\,\textsc{iii}}
\newcommand{\CII}{C\,\textsc{ii}}
\newcommand{\CIV}{C\,\textsc{iv}}
\newcommand{\CrII}{\textrm{Cr}\,\textsc{ii}}

\newcommand{\FeII}{Fe\,\textsc{ii}}
\newcommand{\Ha}{H$\alpha$}
\newcommand{\HI}{\textrm{H}\,\textsc{i}}

\newcommand{\HII}{\textrm{H}\,\textsc{ii}}

\newcommand{\Lya}{Ly$\alpha$}
\newcommand{\lya}{Ly$\alpha$}

\newcommand{\NHI}{$N(\textrm{H}\,\textsc{i})$}

\newcommand{\NI}{N\,\textsc{i}}

\newcommand{\NiII}{Ni\,\textsc{ii}}
\newcommand{\OI}{O\,\textsc{i}}
\newcommand{\SII}{S\,\textsc{ii}}
\newcommand{\SiII}{Si\,\textsc{ii}}
\newcommand{\SiIII}{Si\,\textsc{iii}}
\newcommand{\SiIV}{Si\,\textsc{iv}}
\newcommand{\TiII}{Ti\,\textsc{ii}}
\newcommand{\ZnII}{\textrm{Zn}\,\textsc{ii}}

\def\ltsima{$\; \buildrel < \over \sim \;$}
\def\simlt{\lower.5ex\hbox{\ltsima}}
\def\gtsima{$\; \buildrel > \over \sim \;$}
\def\simgt{\lower.5ex\hbox{\gtsima}}

\title[A DLA and associated Ly$\alpha$ emission in UM\,673A,B]
{A newly discovered DLA and associated Ly$\alpha$ emission in the spectra of 
the gravitationally lensed quasar UM\,673A,B\thanks{Based on 
data obtained at the W.~M. Keck Observatory, 
which is operated as a scientific partnership among the 
California Institute of Technology, the University of California, and 
NASA, and was made possible by the generous financial support of the W.~M. Keck Foundation.}}

\author[Cooke et al.]{Ryan Cooke$^{1}$\thanks{email: rcooke@ast.cam.ac.uk}, 
Max Pettini$^{1,2}$, Charles C. Steidel$^3$, Lindsay J. King$^1$, Gwen C. Rudie$^3$,
\newauthor and Olivera Rakic$^4$\\
$^1$Institute of Astronomy, Madingley Road, Cambridge, CB3 0HA\\ 
$^2$International Centre for Radio Astronomy Research,
        University of Western Australia, 35 Stirling Highway, Crawley,
        WA 6009, Australia\\
$^3$California Institute of Technology, MS 249-17, Pasadena, CA 91125, USA\\
$^4$ Leiden Observatory, Leiden University, P.O. Box 9513, 2300 RA, Leiden, The Netherlands}

\begin{document}

\date{Accepted . Received ; in original form }
\pagerange{\pageref{firstpage}--\pageref{lastpage}} 
\pubyear{2010}

\maketitle

\label{firstpage}

\begin{abstract}
The sightline to the brighter member of the gravitationally lensed quasar pair UM\,673A,B
intersects a damped \Lya\ system (DLA) at $z = 1.62650$ which, because of its
low redshift, has not been recognised before. 
Our high quality echelle spectra of the pair, obtained with HIRES
on the Keck~{\sc i} telescope, show a drop in neutral hydrogen 
column density $N$(\HI) by a factor of at least 400 between 
UM\,673A and B, indicating that the DLA's extent in this direction is  
much less than the $2.7 \, h_{70}^{-1}$\,kpc separation between 
the two sightlines at $z = 1.62650$.
By reassessing this new case together with published data on other quasar pairs,
we conclude that the typical size (radius) of DLAs at these redshifts
is  $R \simeq (5 \pm 3) \, h_{70}^{-1}$\,kpc,
smaller than previously realised. Highly ionized gas
associated with the DLA is more extended, as we find 
only small differences in the \CIV\ absorption profiles
between the two sightlines.
  
Coincident with UM\,673B, we detect a weak and narrow \Lya\ 
\emph{emission} line which we attribute to 
star formation activity 
at a rate  ${\rm SFR} \simgt 0.2$\,M$_\odot$\,yr$^{-1}$.
The DLA in UM\,673A is metal-poor, with an overall metallicity
$Z_{\rm DLA} \simeq 1/30 Z_\odot$, and has a very low internal
velocity dispersion. It exhibits some apparent peculiarities in
its detailed chemical composition, with the elements Ti, Ni, and Zn
being deficient relative to Fe by factors of 2--3. The [Zn/Fe]
ratio is lower than those measured in any other DLA or Galactic halo
star, presumably reflecting somewhat unusual previous enrichment
by stellar nucleosynthesis.
We discuss the implications of these results for the nature
of the galaxy hosting the DLA.

\end{abstract}



\begin{figure*}
  \centering
  \includegraphics[angle=0,width=160mm]{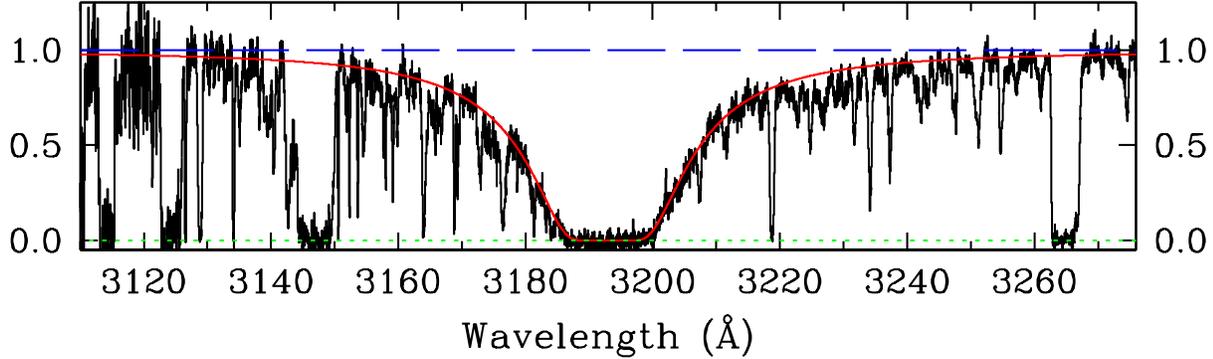}
  \caption{ 
Portion of the HIRES spectrum of UM\,673A (black histogram) encompassing 
the damped \Lya\ line at $z_{\rm abs}=1.626498$. 
The red continuous line shows the theoretical Voigt profile for a neutral
hydrogen column density
$\log[N({\rm H}\,\textsc{i})/{\rm cm}^{-2}]=20.7$. 
The normalised continuum and zero-level are shown by the blue dashed 
and green dotted lines respectively. The $y$-axis scale is residual intensity.
  }
  \label{fig:um673aDLA}
\end{figure*}

\section{Introduction}
UM\,673 (Q0142$-$100) was first identified 
as a gravitationally lensed quasar (QSO) at $z_{\rm em} = 2.719$
by \citet{Sur87} who showed it to be lensed 
by a $z=0.49$ galaxy into two images, UM\,673A and UM\,673B,
separated by 2.2\,arcsec and with magnitudes $m_{\rm R} = 16.9$ and 19.1 respectively.
This pair was later examined in detail by \citet{Sme92}
in their study of intergalactic \Lya\ forest absorbers. 
By considering the number of absorption lines in common (and not in common) 
between the two closely spaced sightlines, these authors were able
to place interesting limits on the typical sizes of \Lya\ clouds, 
following on from analogous analyses by \citet{Sar82} and \citet{Fol84}.

Since these early papers, there have been many
investigations of the \Lya\ forest in the spectra  of gravitationally lensed QSOs.
However, very little is still known about the size and geometry 
of the neutral gas reservoirs that give rise to DLAs with hydrogen column densities
\NHI\,$\geq\,2\times10^{20}\,{\rm cm}^{-2}$, since
these are much rarer absorption systems and thus unlikely to be 
found in front of gravitationally lensed QSOs which are themselves unusual
alignments. 
A recent study by \citet{MonTurRao09} of 
the quadruply lensed Cloverleaf QSO (H\,1413+117) 
with the \textit{Hubble Space Telescope} uncovered
three new DLAs or sub-DLAs at $z \sim 1.5$, none of which
are common to all four components. 
When considered together with analogous
observations of four other DLAs from the literature,
these data led \citet{MonTurRao09} to conclude
that absorbers at 
$z \sim 1.5$ with \NHI\,$=(6-13)\times10^{20}\,{\rm cm}^{-2}$
have typical scale-lengths
${\rm S}_{\rm DLA}=(6-12)\,h_{70}^{-1}\,{\rm kpc}$
(where $h_{70}$ is the Hubble constant in units 
of 70\,km~s$^{-1}$~Mpc$^{-1}$).

Apparently at odds with this conclusion is the finding
by \citet{Ell07} of coincident damped Lyman alpha absorption
on 100\,kpc-scale towards the binary QSO SDSS~1116+4118A,B.
As these dimensions far exceed those expected for a single
galaxy, \citet{Ell07} favour an explanation in terms of a group
of two or more galaxies intersected along these lines of sight.

In this paper we report the discovery of a previously unrecognised
DLA, at $z_{\rm abs} = 1.62650$, in the spectrum of UM\,673A
from high resolution and high signal-to-noise ratio (S/N) spectroscopy
with the HIRES instrument on the Keck\,\textsc{i} telescope.
In line with the compilation of similar
measurements by \citet{MonTurRao09}, no DLA is seen in front of
UM\,673B, even though at the redshift of the DLA the two
sightlines are separated by less than 3\,kpc. 

We do, however, find a weak \Lya\ \emph{emission} line in the 
spectrum of UM\,673B at the same redshift as the DLA.
There have been only a few reported cases of \Lya\ emission
associated with a DLA since the 
first such detection by \citet{HunPetFle90},
as summarised in the review of DLA properties by
\citet{WolGawPro05} with more recent 
updates by \citet{Kul06} and \citet{Chr07}.
Any new examples are of interest in view 
of the apparent puzzle presented by the lack of
obvious star formation associated with gas-rich DLAs
\citep{WolChe06}, and the recent claim that a newly
discovered population of faint line emitters represents
the long-sought host galaxies of DLAs \citep{Rau08}.

Finally, the newly discovered DLA in UM\,673A is metal-poor,
with metallicity $Z \sim 1/30 Z_\odot$. 
Such chemically unevolved DLAs are important,
in that they can provide
clues to early episodes of metal enrichment
in the Universe, complementing efforts being directed
to analogous studies of metal-poor stars in the Milky Way
and nearby dwarf galaxies \citep[see, for example,][]{Pet06}.
In the present case, we
measure the relative abundances of nine different 
elements, from N to Zn,  
and uncover some chemical peculiarities 
which have not been noted before.

The paper is organised as follows.
In Section~\ref{sec:UM673_obs}, we briefly describe 
the observations of UM\,673A,B and the reduction of the HIRES spectra.
In the subsequent analysis, we first focus on the
H\,\textsc{i} gas in front of the UM\,673 pair
(Section~\ref{sec:DLA}), and consider the data presented here
together with the compilation by \citet{MonTurRao09}
of other DLAs in gravitationally
lensed QSOs to refine those authors'
estimate of the characteristic size of damped systems (Section~\ref{sec:sizes}).
We next turn to the \Lya\ emission detected in the spectrum of
UM\,673B at the same redshift as the DLA in UM\,673A (Section~\ref{sec:lya}),
and use it to obtain an estimate of the star formation rate in
the galaxy associated with the DLA.
Section~\ref{sec:metals} deals with the chemical
composition of the DLA, comparing it to that of
DLAs and Galactic halo stars of similar overall metallicity.
We summarise our findings and draw some conclusions
in Section~\ref{sec:conc}.
Throughout the paper, we adopt a `737' cosmology, with $
H_0=70\,{\rm km\,s}^{-1}\,{\rm Mpc}^{-1}$, 
$\Omega_{\rm M}=0.3$ and $\Omega_{\Lambda}=0.7$.

\section{Observations and Data Reduction}
\label{sec:UM673_obs}

We observed UM\,673A and UM\,673B 
on the nights of 2005 October 9 and 10, 
and again three years later on the nights of
2008 September 24 and 25,
as part of a large-scale imaging and spectroscopic survey
of galaxies in the fields of bright QSOs, aimed
primarily at studying the outflows of interstellar 
gas from star-forming galaxies at redshifts $z = 2$--3 \citep{Ade05,Ste10}.
HIRES \citep{Vog94} was configured to cover the 
wavelength range 3100--6100\,\AA\ (with small gaps
near 4000\,\AA\ and 5000\,\AA\ due to gaps between the 
three CCD chips on the detector) using the ultraviolet (UV) cross-disperser
and collimator. 

In order to avoid cross-contamination between the two images
and to minimize slit losses due to atmospheric dispersion,
UM\,673 A and B were observed separately, with the HIRES
slit maintained at the parallactic angle by its image rotator.
For UM\,673\,A we used a 1.15\,arcsec-wide entrance slit,
which results in a resolution $R \equiv \lambda/\Delta \lambda = 36\,000$,
corresponding to a velocity full width at half maximum
${\rm FWHM} = 8.3$\,km~s$^{-1}$, sampled with
$\sim 3$ pixels. 
The total integration time was  $9400\,{\rm s}$, divided into five
exposures; the QSO was stepped along the slit between
each exposure. 
For the fainter UM\,673\,B, we employed the narrower
0.86\,arcsec slit (so as to exclude more effectively light from the 
brighter image) which results in $R = 48\,000$ and 
${\rm FWHM} = 6.2$\,km~s$^{-1}$ sampled with $\sim 2$ detector pixels.
The total exposure time was $28\,200\,{\rm s}$, again
divided into a number of separate exposures, typically 2700\,s long.
The seeing was $\le 1 $\,arcsec FWHM throughout
the observations. 

To these data we added another set of observations of UM\,673A,B,
which we retrieved from the Keck Observatory data archive, 
obtained in 1996 with the original HIRES red-sensitive detector
and red-optimized cross-disperser. While these earlier data
do not contribute much at blue and ultraviolet wavelengths,
with their long exposure times (18\,000\,s and 27\,000\,s for 
UM\,673A and B respectively) they do improve the S/N ratio 
of our final co-added spectrum at red wavelengths.

The two-dimensional HIRES spectra were processed with the
\textsc{makee} data reduction pipeline developed by Tom Barlow
which includes the usual steps of flat-fielding, order tracing, background
subtraction, 1-D extraction and merging of the echelle orders.
A wavelength reference was provided by the spectrum of the
internal Th-Ar hollow cathode lamp and the co-added,
1-D spectra were mapped onto a vacuum heliocentric wavelength scale.
In a final step, the spectra were normalised by dividing out the
QSO continuum and emission lines. The rms deviations of the 
data from the continuum in regions free from absorption lines
provide an empirical measure of the signal-to-noise ratio.
For UM\,673A, our data have ${\rm S/N} >  35$ per
2.7\,km~s$^{-1}$ ($\approx 0.04$\,\AA) bin from
$\sim 4000$\,\AA\ to $\sim 6000$\,\AA; the S/N is highest
near 5000\,\AA\ (${\rm S/N} \simeq 70$) and is still moderately
high (${\rm S/N} \simeq 24$) at 3200\,\AA,
near the redshifted wavelength of the damped \Lya\ line.
The corresponding values for UM\,673B are 
${\rm S/N} \simgt 15$ (4000--6000\,\AA) and 
${\rm S/N} \simeq 8$ at 3200\,\AA.

\section{H\,\textsc{i} Absorption towards UM\,673A,B}
\label{sec:DLA}

\subsection{The DLA towards UM\,673A}

Our HIRES spectra extend to shorter wavelengths than most previous
observations of this famous QSO pair, which probably explains why the
damped \Lya\ system in front of UM\,673A 
(see Fig.~\ref{fig:um673aDLA})
has gone unnoticed until now. 
Associated with the DLA are a multitude
of metal absorption lines of elements from C to Zn. 
These lines are analysed in detail in Section~\ref{sec:metals};
for the present purpose suffice it to say that they are
narrow, with ${\rm FWHM} \simlt 25$\,km~s$^{-1}$,
and have maximum optical depth at $z_{\rm abs} = 1.626498$.
Adopting the same redshift for the damped \Lya\ line
(which is too broad to allow such an accurate  determination of
$z_{\rm abs}$), we find  
$\log[N({\rm H}\,\textsc{i})/{\rm cm}^{-2}]=20.7 \pm 0.1$
by fitting theoretical Voigt profiles to the wings 
of the line (and interpolating across narrower absorption
features---see Fig.~\ref{fig:um673aDLA}).


\begin{figure}
  \centering
  \includegraphics[width=80mm]{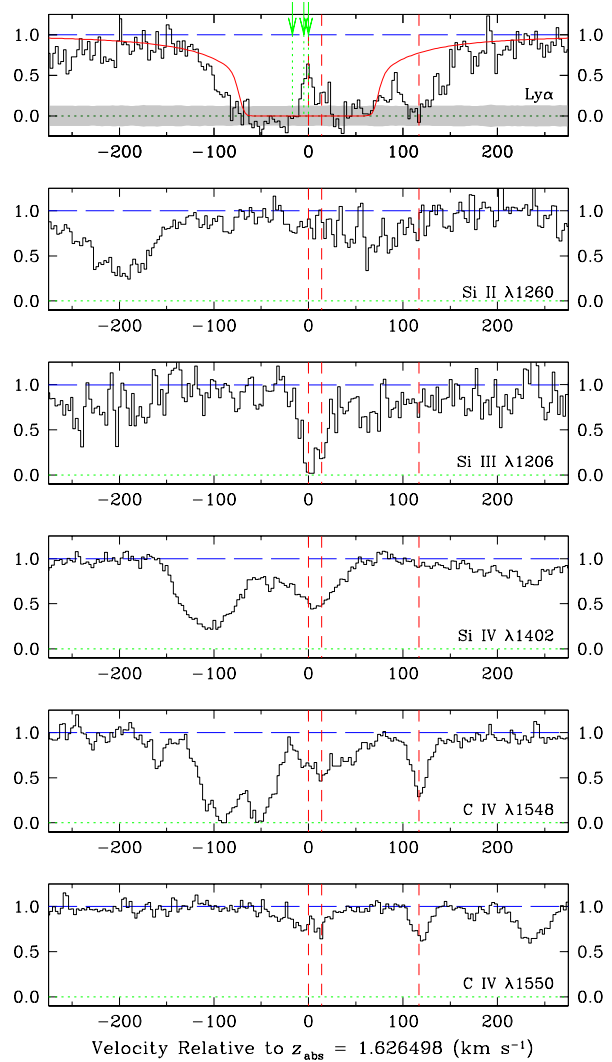}
  \caption{
A selection of absorption lines in UM\,673B near the redshift of
the DLA at $z_{\rm abs}  = 1.626498$ in front of UM\,673A; 
in all panels the $y$-axis is residual intensity.
The three green arrows in the top panel indicate the velocities
of the three components of the low-ionization metal lines in the 
DLA (see Section~\ref{sec:metals}), 
while the three long-dash red lines
through all the panels mark the velocities of absorption (and
emission) components in UM\,673B.
In the top panel, the red continuous line is the 
theoretical \Lya\ absorption profile
for the \textit{upper limit} we deduce to the column density 
of neutral hydrogen in UM\,673B, 
$\log[$\NHI$/{\rm cm}^{-2}]=18.1$.
The grey shaded area shows the $\pm 1 \sigma$ error spectrum.
Note the detection of \Lya\ \textit{emission} in the core
of the strong absorption line, at  the same redshift as the DLA
in UM\,673A. The emission profile also aligns well with
the high ionization absorption lines in UM\,673B reproduced 
in the lower four panels.
 }
   \label{fig:um673blya}
\end{figure}

\subsection{The Lyman limit system in UM\,673B}

We also find absorption near $z_{\rm abs} = 1.626498$
in the spectrum of UM\,673B, but with much reduced column
densities of neutral gas.  
The top panel of Fig.~\ref{fig:um673blya} shows the
wavelength region around the \Lya\ line which is
broad and saturated, but not damped.
Under these circumstances, it is well known that
the column density  is unconstrained within orders of magnitude, 
unless higher order Lyman lines are available---this is the reason
why the H\,\textsc{i} column density distribution
is so poorly sampled in the interval 
$\log[$\NHI$/{\rm cm}^{-2}] = 17$--20 \citep[e.g.][]{Sto00,Ome07}.

However, since \NHI\ and the velocity dispersion parameter $b$ (km~s$^{-1}$)
are degenerate in strongly saturated lines, we can still determine
an upper limit to the column density by considering the smallest
$b$-value, and corresponding highest value of \NHI, which 
provide an acceptable fit to the width and profile of the
saturated \Lya\ line in UM\,673B.
To this end, we considered a series of pair values of $b$ and \NHI,
fixing \NHI\ and using \textsc{vpfit}\footnote{\textsc{vpfit} is available 
from http://www.ast.cam.ac.uk/${\sim}$rfc/vpfit.html} to determine the value of $b$
for which the theoretical line profile shows the least disagreement
with the data.

We started the iteration at 
$\log[$\NHI$/{\rm cm}^{-2}] = 20.7$, as measured
in UM\,673A and which greatly overproduces the
observed \Lya\ absorption in UM\,673B for all values of
$b$, and then decreased $\log$\,\NHI\ in steps of 0.1, until a
plausible fit was arrived at
for $\log[$\NHI$/{\rm cm}^{-2}] = 18.1$ and $b=22\,{\rm km\,s}^{-1}$.
The corresponding line profile, convolved with the instrumental
resolution, is superimposed on the data in the top panel of  
Fig.~\ref{fig:um673blya}.
While the absorption in the line core is less than observed,
presumably because of neighbouring \Lya\ absorption lines---one 
of which, at $\Delta v = +114$\,km~s$^{-1}$, 
is incidentally also seen as a redshifted
component in \CIV---higher column densities of \NHI\ would overproduce
the absorption in the line wings, relative to what is observed.
We consider $\log[$\NHI$/{\rm cm}^{-2}] \le 18.1$ to be an
upper limit to the column density of neutral gas in UM\,673B
because equally good or even better fits could be obtained
with lower values of \NHI\ and higher values of $b$.

Thus, we are led to the conclusion that the column density of neutral hydrogen
drops by a factor of at least 400 over a transverse distance of less than 3\,kpc
(Section~\ref{sec:sizes}). A comparable drop is deduced from consideration
of metal absorption lines from ionization stages which are dominant
in H\,\textsc{i} regions. For example, in Section~\ref{sec:metals},
we deduce a column density $\log[N({\rm Si}\,\textsc{ii})/{\rm cm}^{-2}]=14.75 \pm 0.03$
from the analysis of five Si\,\textsc{ii} transitions in UM\,673A. The strongest of these,
Si\,\textsc{ii}~$\lambda 1260.4221$, is below the detection limit in
UM\,673B (see second panel from the top in Fig.~\ref{fig:um673blya}).
In the optically thin limit,
\begin{equation}
\label{eqn:ew}
N=1.13 \times 10^{20} \cdot
\frac{W_\lambda}{\lambda^2 f} 
~~ {\rm cm}^{-2} 
\end{equation}
where $W_\lambda$ and $\lambda$ are respectively
the rest frame equivalent width
and wavelength (both in \AA), and $f$
is the oscillator strength.\footnote{Throughout
this work, we use the compilation of laboratory
wavelengths and $f$-values by \citet{Mor03}
with updates by \citet{Jen06}.}
From equation~(\ref{eqn:ew})
we deduce $\log[N({\rm Si}\,\textsc{ii})/{\rm cm}^{-2}] \leq  12.2$
($3 \sigma$ limit)
in UM\,673B, a factor of $\geq 350$ lower than in 
UM\,673A.


\begin{figure}
  \centering
  \includegraphics[width=95mm]{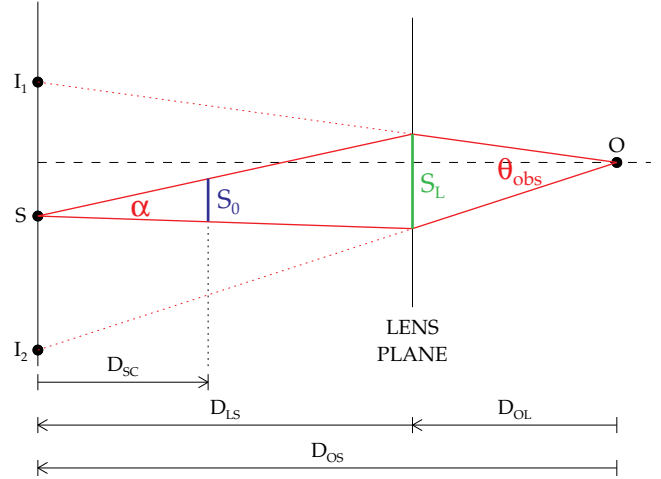}
  \caption{
   The geometry of a gravitational lens, as viewed by the observer at O, 
   with source (S) being lensed into two images (${\rm I_1}$ and ${\rm I_2}$) 
   of angular separation $\theta_{\rm obs}$, 
   by a galaxy situated at the lens plane along the optical axis (dashed horizontal line).
   S$_0$ is the transverse distance between the two sightlines
   at the location of the galaxy producing the DLA.
   }
   \label{fig:trans_dist}
\end{figure}

\section{Constraining the sizes of DLAs}
\label{sec:sizes}

\begin{table*}
\centering
    \caption{\textsc{Derived e-folding scale lengths of DLAs, updated from Table~5 of \citet{MonTurRao09}}}
    \begin{tabular}{@{}llllccccccc}
    \hline
  \multicolumn{1}{c}{QSO~~~~~~~~~~~}
& \multicolumn{1}{c}{$z_{\rm em}$~~}
& \multicolumn{1}{c}{$z_{\rm lens}$~~}
& \multicolumn{1}{c}{$z_{\rm abs}$~~}
& \multicolumn{1}{c}{~~Pair~~}
& \multicolumn{1}{c}{$\theta_{\rm obs}$}
& \multicolumn{1}{c}{$N({\rm H}\,\textsc{i})_{\rm max}$}
& \multicolumn{1}{c}{$N({\rm H}\,\textsc{i})_{\rm min}$}
& \multicolumn{1}{c}{${\rm S}_{0}^{\rm a}$}
& \multicolumn{1}{c}{${\rm S}_{\rm DLA,\bar{e}}^{\rm b}$}
& \multicolumn{1}{c}{${\rm S}_{\rm DLA,e}^{\rm c}$}\\
    \multicolumn{1}{c}{}
& \multicolumn{1}{c}{}
& \multicolumn{1}{c}{}
& \multicolumn{1}{c}{}
& \multicolumn{1}{c}{}
& \multicolumn{1}{c}{(arcsec)}
& \multicolumn{1}{c}{($10^{20}\,{\rm cm}^{-2}$)}
& \multicolumn{1}{c}{($10^{20}\,{\rm cm}^{-2}$)}
& \multicolumn{1}{c}{($h_{70}^{-1}\,{\rm kpc}$)}
& \multicolumn{1}{c}{($h_{70}^{-1}\,{\rm kpc}$)}
& \multicolumn{1}{c}{($h_{70}^{-1}\,{\rm kpc}$)}\\
    \hline
H\,1413+117  & 2.55      & 1.0      & 1.440  & B-A & 0.753 & 60   & 9.0    & 3.13     & 2.33  & 1.65\\
                      &             &             &            & B-D & 0.967 & 60   & 0.25   & 4.02    & 2.55  & 0.733\\
                       &              &             &            & B-C & 1.359 & 60   & 0.20   & 5.65    & 3.58  & 0.991\\
                       &              &            &             & A-D & 1.118 & 9.0  & 0.25   & 4.64     & 3.02  & 1.29\\
                       &              &            &             & A-C & 0.872 & 9.0  & 0.20   & 3.62    & 2.34  & 0.951\\
                       &              &            &            & D-C & 0.893 & 0.25 & 0.20   & 3.71     & 11.7  & 16.6\\
H\,1413+117   &             &             & 1.486  & D-A & 1.118 & 2.0  & \textless0.05  & 4.32    & \textless2.80 & \textless1.17\\
                      &             &              &           & D-B & 0.967 & 2.0  & \textless0.1   & 3.73  &    \textless2.48  &\textless1.25 \\
                     &              &              &           & D-C & 0.893 & 2.0  & \textless0.05  & 3.45  &   \textless2.24  &\textless0.935\\
H\,1413+117   &           &               & 1.662  & B-A & 0.753 & 6.0  & 1.5    & 2.17     & 1.83  & 1.57\\
                     &        & & & B-C & 1.359 & 6.0  & 0.6    & 3.91     & 2.75  & 1.70\\
                     &        & & & B-D & 0.967 & 6.0  & 0.3    & 2.78    & 1.85   & 0.928\\
                     &        & & & A-C & 0.872 & 1.5  & 0.6    & 2.51     & 2.64  & 2.74\\
                     &        & & & A-D & 1.118 & 1.5  & 0.3    & 3.22     & 2.54  & 2.00\\
                     &        & & & C-D & 0.893 & 0.6  & 0.3    & 2.57     & 3.25  & 3.71\\
HE\,0512$-$3329 &1.58      & 0.93      & 0.9313 & A-B & 0.644 & 3.09 & 2.95   & 5.05     & 70.4  & 109  \\
Q\,0957+561    & 1.4136  & 0.36      & 1.3911 & A-B & 6.2   & 1.9  & 0.8    & 0.278   & 0.303 & 0.321\\
UM\,673          & 2.7313  & 0.493    & 1.6265 & A-B & 2.22  & 5.0  & \textless0.013 & 2.71   & \textless1.72 & \textless0.455\\
HE\,1104$-$1805  & 2.31    & 0.73      & 1.6616 & A-B & 3.0   & 6.3  & \textless0.037 & 4.47   & \textless2.84 & \textless0.870\\
    \hline
    \end{tabular}
    \smallskip
\begin{flushleft}
$^{\rm a}$ Transverse separation between the two sightlines at the redshift of the absorber.\\
$^{\rm b}$ $e$-folding scale length of DLA assuming a linear decline of $N$(\HI)---see 
equation~(\ref{eqn:dla_sizes_ebar}).\\
$^{\rm c}$ $e$-folding scale length of DLA assuming an exponential decline of $N$(\HI)---see 
equation~(\ref{eqn:dla_sizes_e}).\\
\end{flushleft}
   \label{tab:dla_sizes}

\end{table*}

In this section we use the finding that the DLA in UM\,673A 
is not present in the spectrum of UM\,673B to reassess, together
with existing data, the characteristic size of
the \HI\ clouds giving rise to damped \Lya\ systems.

We begin with a simple derivation of the transverse distance, 
similar to that presented by \citet{Sme92}. 
Referring to Fig.~\ref{fig:trans_dist},
the transverse distance between the two images at the lens plane is 
${\rm S_{L}}= \theta_{\rm obs} \, {\rm D_{OL}}=\alpha \, {\rm D_{SL}}$, 
and the transverse distance between the two light paths at the redshift 
of the absorber is ${\rm S_{0}}=\alpha \, {\rm D_{SC}}$,
where ${\rm D_{OL}}$, ${\rm D_{SL}}$, ${\rm D_{SC}}$ 
are the angular diameter distances from, respectively,
the observer to the lens, the source to the lens, and the
source to the absorbing cloud.

Thus, the transverse distance between the light paths
at the redshift of the absorber can be written as
\begin{equation}
{\rm S_{0}}=\frac{\theta_{\rm obs} \, {\rm D_{OL} \, D_{SC}}}{\rm D_{SL}}
= \frac{\theta_{\rm obs} \, {\rm D_{OL} \, D_{CS}}}{\rm D_{LS}}
\cdot
\frac{(1+z_{\rm L})}{(1+z_{\rm C})}
\label{eq:trans_cloud_01}
\end{equation}
where $z_{\rm L}$ and $z_{\rm C}$ are the redshifts of the lens and 
the DLA respectively. 
Recalling that the angular diameter distance 
between two objects at redshift $z_{2}$ and $z_{1}$, where $z_{2}\,>\,z_{1}$, 
is of the form \citep{Hog99},
\begin{equation}
{\rm D_{12}}=\frac{{\rm D}_{2} - {\rm D}_{1}}{1+z_{2}},
\end{equation}
where
\begin{equation}\label{eq:com_dist}
{\rm D}_{i} = \frac{c}{H_{0}}\int_{0}^{z_{i}}\frac{dz}{\sqrt{\Omega_{\Lambda}+(1+z)^{3}\Omega_{\rm M}}}
\end{equation}
is the comoving distance to redshift $z_{i}$,
we can rewrite equation~(\ref{eq:trans_cloud_01}) as:
\begin{equation}
{\rm S_{0}}=\frac{\theta_{\rm obs} \, {\rm D_{L} \, (D_{S}-D_{C})}}{(1+z_{\rm C}){\rm (D_{S}-D_{L})}} \, .
\end{equation}
Thus, adopting $\theta_{\rm obs}=2.22$\,arcsec, 
$z_{\rm L}=0.493$ \citep{Sur88,Leh00}, 
$z_{\rm S}=2.7434$ from our own unpublished near-infrared observations
of the H$\beta$ emission line\footnote{As is normally the case,
the systemic redshift deduced from the Balmer lines is higher
than the redshift indicated by the rest-frame ultraviolet emission lines,
in this case $z_{\rm em} = 2.719$  from the EFOSC spectra obtained
by \citet{Sur87} and $z_{\rm em} = 2.7313$ from the Sloan
Digital Sky Survey \citep{Sch07}. The difference
from the original value reported by
\citet{Sur87} is nearly +2000\,km~s$^{-1}$.}, and 
$z_{\rm C}=1.62650$ from the HIRES observations presented here, 
we find that the transverse (physical) distance between the two sightlines 
at the redshift of the DLA is ${\rm S_{0}} =  2.7\,{\rm kpc}$
(in our `737' cosmology).

We now add this new case to the list compiled 
by \citet{MonTurRao09} and reanalyse the entire sample in Table~\ref{tab:dla_sizes}.
Note that we have revised the values for the Cloverleaf (H\,1413+117)
from Table\,5 of \citet{MonTurRao09} for two reasons.
First, in calculating the transverse distances applicable to the two
low redshift systems (at $z_{\rm abs} = 1.440$ and 1.486) in the Cloverleaf,
these authors assumed $z_{\rm L}=z_{\rm C}$ 
(E. Monier, private communication). 
Second, recent mid-infrared data have improved our understanding of this system 
\citep{MacKocAgo09}. 
The positions and relative fluxes of the images can be well explained 
by a lensing galaxy at $z_{\rm L}\approx1.0$ \citep*{KneAllPel98}, 
with an additional galaxy located $\Delta\alpha_{\rm G2}= -1.87$\,arcsec, 
$\Delta\delta_{\rm G2}=4.14$\,arcsec  from the lens, 
coincident with a galaxy identified by \citet{KneAllPel98} as object No. 14. 
The inclusion of this additional galaxy in the lensing model 
does not affect the astrometry (but does affect the relative fluxes of the QSO images);
thus, we assumed a single lens geometry at $z=1.0$. 
We have excluded from the \citet{MonTurRao09} sample the binary QSO 
LBQS\,1429$-$0053 because the linear scale probed by that pair,
${\rm S}_0 = 43.2 \,h_{70}^{-1}$\,kpc, is 
one order of magnitude larger than those of all the other 
(gravitationally lensed) pairs
considered here. As argued by \citet{Ell07},
such large scales are more likely to  probe 
the clustering properties of DLAs rather than the 
typical sizes of their host galaxies.
In any case, our conclusions below on the median size
of DLAs are unaltered by the inclusion or omission of 
LBQS\,1429$-$0053.

In order to determine the characteristic size of DLAs, ${\rm S_{DLA}}$, 
we adopt two simple models. 
In the first, ${\rm S_{DLA}}$ depends linearly on the \HI\ column density, 
while in the second  ${\rm S_{DLA}}$ depends linearly on the 
logarithm of \NHI. 
The latter is probably more realistic, 
as it corresponds to an exponential decline of \NHI, 
but we also 
consider the former for comparison 
with the analysis by \citet{MonTurRao09}. 
Our analysis, however, differs from theirs in the following way.
\citet{MonTurRao09} examined \NHI\ as a function of the 
observed \emph{angular} separation of two QSO images, $\theta_{\rm obs}$, 
whereas we considered \NHI\ as a function of the 
\textit{physical} (transverse) distance between 
the two sightlines at the redshift of the absorber.
The approach taken by \citet{MonTurRao09} has the advantage
of being independent of the choice of cosmological parameters,
while ours is perhaps more physically motivated.
We therefore have an equation of the form,

\begin{equation} 
{\rm S}={\rm S}_{0} \,
\frac{N({\rm H}\,\textsc{i})_{\rm max}-N({\rm H}\,\textsc{i})}
{N({\rm H}\,\textsc{i})_{\rm max}-N({\rm H}\,\textsc{i})_{\rm min}},
\end{equation}
where $N({\rm H}\,\textsc{i})_{\rm max}$ is 
the higher column density observed between any two given sightlines, 
and $N({\rm H}\,\textsc{i})_{\rm min}$ is the lower column density of the two. 
In the first  model, we follow \citet{MonTurRao09} and 
define the linear $e$-folding scale-length 
(${\rm S}_{\rm DLA,\bar{e}}$) to be the transverse distance over which 
\NHI$_{\rm max}$ decreases by a factor of $e=2.718$ 
(i.e. $N({\rm H}\,\textsc{i})=N({\rm H}\,\textsc{i})_{\rm max}/2.718$),
\begin{equation}
\label{eqn:dla_sizes_ebar}
{\rm S}_{\rm DLA,\bar{e}} = \,{\rm S}_{0}\,
\frac{0.632 \, N({\rm H}\,\textsc{i})_{\rm max}}
{N({\rm H}\,\textsc{i})_{\rm max}-N({\rm H}\,\textsc{i})_{\rm min}}.
\end{equation}
Note that one can 
easily convert to the `e-folding angle', 
$\theta_{\rm e}$ introduced by \citet{MonTurRao09} using
the relation: 
$\theta_{\rm e} = \theta_{\rm obs}\,{\rm S_{DLA,\bar{e}}}/{\rm S_{0}}$.

In the second case considered,  ${\rm S}$ scales with $\ln$\,\NHI:
\begin{equation}
N({\rm H}\,\textsc{i})=N({\rm H}\,\textsc{i})_{\rm max}\,\exp[-{\rm S}/{\rm S}_{\rm DLA,e}] \, .
\label{eqn:exp}
\end{equation}
We can then define the \emph{true} $e$-folding scale length for DLAs,
\begin{equation}
\label{eqn:dla_sizes_e}
{\rm S}_{\rm DLA,e} =
\frac{{\rm S}_{0}}{
{\rm ln}[N({\rm H}\,\textsc{i})_{\rm max}/N({\rm H}\,\textsc{i})_{\rm min}]},
\end{equation}
where ${\rm S}_{0}$, $N({\rm H}\,\textsc{i})_{\rm max}$ and $N({\rm H}\,\textsc{i})_{\rm min}$ all take their previous definitions. 

\begin{figure}
  \centering
  \includegraphics[angle=0,width=80mm]{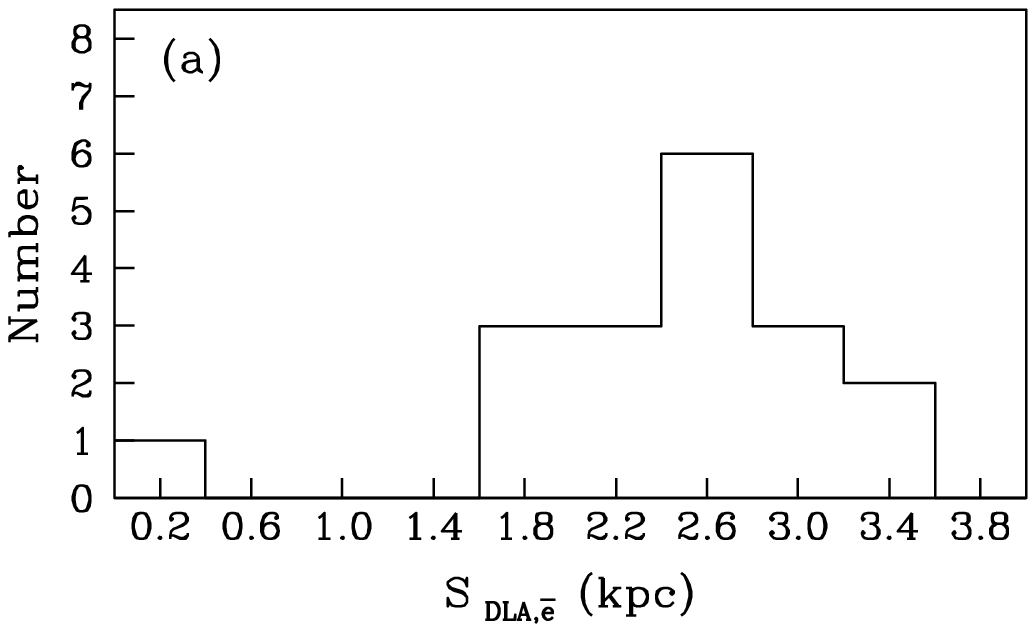}
  \includegraphics[angle=0,width=80mm]{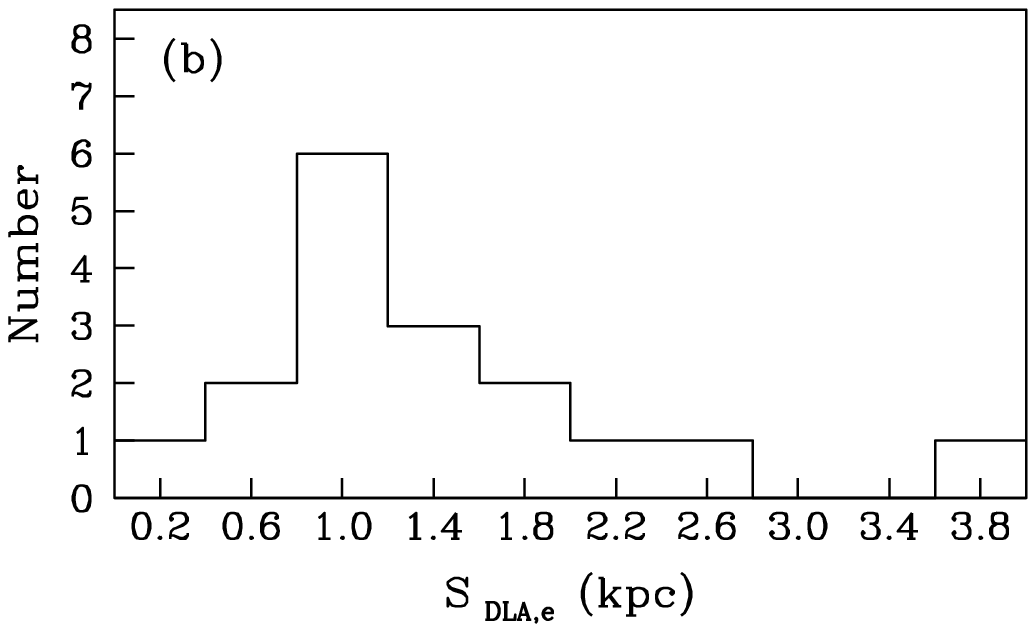}
  \caption{
  The distributions of (a) linear $e$-folding scale lengths and 
  (b) $e$-folding scale lengths for DLAs, based on 
  equations~(\ref{eqn:dla_sizes_ebar}) and (\ref{eqn:dla_sizes_e}) respectively.
   Upper limits in the values of 
   $\tilde{\rm S}_{\rm DLA,\bar{e}}$ 
   and $\tilde{\rm S}_{\rm DLA,e}$
   in Table~\ref{tab:dla_sizes} have been plotted as if they were measured values.
   For simplicity, two outlying values in excess of 10\,kpc have been omitted from the
   plots.
   }
   \label{fig:dla_sizes}
\end{figure}

In both models,  the $e$-folding scale lengths 
are most uncertain when 
$N({\rm H}\,\textsc{i})_{\rm max}/N({\rm H}\,\textsc{i})_{\rm min}\approx 1$, 
due to the large extrapolation required. 
The derived $e$-folding scale lengths for each DLA 
are listed in Table~\ref{tab:dla_sizes}, 
and the two distributions are shown with histograms
in Fig.~\ref{fig:dla_sizes}. 
Considering all the measurements, 
we determine the median linear $e$-folding 
and median $e$-folding scale lengths of DLAs 
to be $\tilde{\rm S}_{\rm DLA,\bar{e}}=2.6 \pm 0.7\,{\rm kpc}$ 
and $\tilde{\rm S}_{\rm DLA,e}=1.3 \pm 0.8\,{\rm kpc}$ respectively.
The errors were computed with the Interactive Data Language routine
\textsc{robust\_sigma}\footnote{Available 
from http://idlastro.gsfc.nasa.gov/homepage.html}  
which determines the median absolute 
deviation (unaffected by outliers) of a set of measurements, 
and then appropriately weights the data to provide a
robust estimate of the sample dispersion \citep{HoaMosTuk83}.

The values of $\tilde{\rm S}_{\rm DLA,\bar{e}}$ and
$\tilde{\rm S}_{\rm DLA,e}$ we deduce
are lower than that reported by \citet{MonTurRao09},
${\rm S}_{\rm DLA}=6\,h_{70}^{-1}\,{\rm kpc}$,
partly because of the 
improved estimate of the lens redshift in the Cloverleaf.
Indeed, given the large number of sightline pairs in this multiple system
(see Table~\ref{tab:dla_sizes}), the uncertainty in the lens redshift
of the Cloverleaf still has a marked effect on the values
of $\tilde{\rm S}_{\rm DLA,\bar{e}}$ and $\tilde{\rm S}_{\rm DLA,e}$
deduced. In order to assess the effect quantitatively,
we have repeated the above analysis with the rather extreme
assumptions that the lens redshift is, in turn, 
$z_{\rm L} = 0.5$ and 1.5, instead of
the value $z_{\rm L} = 1.0$ adopted in Table~\ref{tab:dla_sizes}
from \citet{KneAllPel98}.
We find 
$\tilde{\rm S}_{\rm DLA,\bar{e}}=1.0 \pm 0.3\,{\rm kpc}$, 
$\tilde{\rm S}_{\rm DLA,e}=0.5 \pm 0.3\,{\rm kpc}$
if $z_{\rm L} = 0.5$,
and
$\tilde{\rm S}_{\rm DLA,\bar{e}}=5.6 \pm 1.9\,{\rm kpc}$, 
$\tilde{\rm S}_{\rm DLA,e}=2.7 \pm 2.0\,{\rm kpc}$
if $z_{\rm L} = 1.5$. 
The range of values of S$_{\rm DLA}$ admitted by 
the data will narrow as more QSO pairs are studied
in the future.

Finally, we stress that our estimates of S$_{\rm DLA}$ are 
not the same as what is generally thought of as the
`size' of a DLA. Referring to equation~(\ref{eqn:exp}),
if we assume an idealised spherical cloud with peak
$N$(H\,\textsc{i})$_{\rm max} = 1 \times 10^{22}$\,cm$^{-2}$
at its centre,
we have to move a radial distance ${\rm S} = \ln 50 \, {\rm S}_{\rm DLA, e}$,
or ${\rm S} \simeq 4 \, {\rm S}_{\rm DLA, e}$,
before $N$(H\,\textsc{i}) falls below the
threshold
$N$(H\,\textsc{i})$_{\rm min} = 2 \times 10^{20}$\,cm$^{-2}$ 
generally adopted  
as the definition of a DLA.
Thus, our preferred solution,
$\tilde{\rm S}_{\rm DLA,e}=1.3 \pm 0.8\,{\rm kpc}$,
corresponds to DLA radii of $\sim 5 \pm 3$\,kpc.


\section{\Lya\ emission towards UM\,673B}
\label{sec:lya}

Returning to Fig.~\ref{fig:um673blya} (top panel), the presence
of a weak emission line in the core of the saturated \Lya\ absorption
line can be readily recognised. 
Although in principle this feature could also be a gap between
two adjacent absorption lines, its precise wavelength
match with: (a) high ionization metal absorption lines along
the same sightline (lower four panels in 
Fig.~\ref{fig:um673blya}), and (b) the highest
optical depth absorption in the DLA in UM\,673A
argues in favour of our interpretation as a weak and narrow
\Lya\ emission line.

\begin{figure}
  \centering
  \hspace*{-0.2cm}
  \includegraphics[width=86mm]{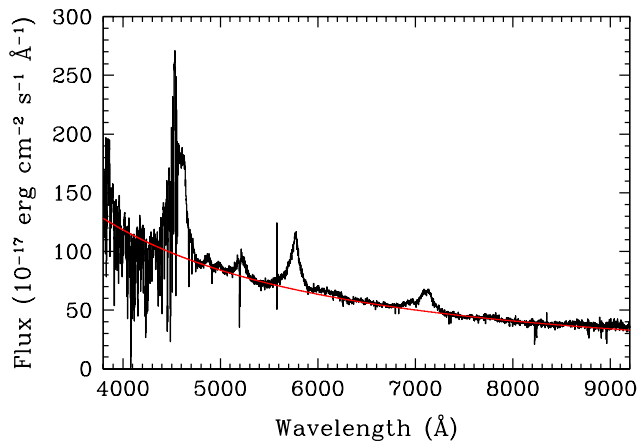}
  \caption{
   SDSS spectrum of UM\,673A,B (black line) together with our power-law fit 
   to the continuum (red line) of the form 
   ${\cal F}_\lambda = A \cdot (\lambda/{\rm \AA})^{-\beta}$, 
   with best fitting values
   $\beta = 1.535$ and 
   $A = 4.0\times10^{-10}\,{\rm erg\,s}^{-1}\,{\rm cm}^{-2}\,{\rm \AA}^{-1}$.
   }
   \label{fig:sdss}
\end{figure}

In order to measure the line flux and luminosity, we referred our echelle
spectrum (for which the absolute flux calibration can be uncertain) 
to the Sloan Digital Sky Survey (SDSS) spectrum of UM\,673A,B,
reproduced in Fig.~\ref{fig:sdss}.
Fitting the continuum longwards of the QSO \Lya\ emission line
with a power-law of the form
${\cal F}_\lambda = A \cdot (\lambda/{\rm \AA})^{-\beta}$,
we deduced the best-fitting values $\beta = 1.535 \pm 0.005$ and
$A = (4.0 \pm 0.2) \times10^{-10}\,{\rm erg\,s}^{-1}\,{\rm cm}^{-2}\,{\rm \AA}^{-1}$
for the slope and normalization respectively.
The fit, which is shown with a red line in 
Fig.~\ref{fig:sdss}, reproduces the sum of the SDSS $r$ magnitudes
of UM\,673A,B ($r = 16.73$ and 18.84, respectively)
when convolved with the transmission curve
of the $r$-band filter.
Extrapolating this continuum to $\lambda_{\rm obs} = 3193$\,\AA,
where the redshifted \Lya\ line at $z = 1.62650$ falls, 
and allowing for the fraction of the light contributed by UM\,673A,
then provides
an absolute flux scale for the continuum shown by the long-dash line
in the top panel of Fig.~\ref{fig:um673blya},
where a residual intensity of  1.00 corresponds to a flux density
$\cal{F}_{B}$($3193$\AA)\,$=(2.1 \pm 0.4) 
\times10^{-16}\,{\rm erg\,s}^{-1}\,{\rm cm}^{-2}\,{\rm \AA}^{-1}$.
The 20\% error is the systematic uncertainty
in the flux calibration due to the 
combined effects of: (i) extrapolation of the QSO
continuum to 3193\,\AA, and (ii) the accuracy of the SDSS photometry. 

Integrating across the \Lya\ emission line then yields a line flux
$F$(Ly$\alpha) = (2.5 \pm 0.25 \pm 0.5) \times 10^{-17}$\,erg\,s$^{-1}$\,cm$^{-2}$,
quoting separately the random error from the counting statistics 
(shown by the shaded region
in the top panel of Fig.~\ref{fig:um673blya}) and 
the systematic uncertainty in the flux calibration.
The corresponding line luminosity in our cosmology is
$L({\rm Ly\alpha}) = (4.3 \pm 0.4 \pm 0.9)\times10^{41}\,{\rm erg\,s}^{-1}$.


\subsection{Origin of the \Lya\ emission}

We now discuss some of the mechanisms that could produce 
the observed \Lya\ emission,
before detailing what we consider to be 
the most plausible interpretation. 
The first source we consider is the metagalactic UV background, 
which could produce \Lya\ emission by fluorescence, 
the so-called Hogan-Weymann effect \citep{HogWey87}. 
This diffuse emission has yet to be observed, 
presumably because of the low surface brightness it produces.
Given the current understanding of this effect, 
at the redshift of the DLA we would expect to observe 
a surface brightness \citep{GouWei96},
\begin{equation}
\mu \simeq 5.7\times10^{-19}
\big( \frac{\eta_{E}}{0.5} \big)
\bigg( \frac{J}{5.1\times10^{-22}} \bigg) \,\,\,
{\rm erg\,s}^{-1}\,{\rm cm}^{-2}\,{\rm arcsec}^{-2},
\label{eq:surf_bright}
\end{equation}
where $\eta_{E}$ represents the efficiency 
with which the incident UV background is re-radiated 
as fluorescent \Lya\ emission and $J$ is the ionizing background 
at the Lyman limit at $z \sim 2$ \citep{Bol05}. 
Through the area of sky covered by our observations
($ 0.86 \times 3$\,arcsec, the latter being the size of the 
aperture used to extract the 1-D spectra from the raw 2-D HIRES images), 
the surface brightness of eq.~(\ref{eq:surf_bright}) would produce
an integrated line flux
$F({\rm Ly\alpha}) \sim 1.5\times10^{-18}\,{\rm erg\,s}^{-1}\,{\rm cm}^{-2}$, 
one order of magnitude lower than the flux recorded.

\Lya\ fluorescence can also be induced by the UV 
radiation from a nearby AGN impinging on the 
DLA \citep[e.g.][]{Ade06}. Following the 
formalism introduced by \citet{Can05}, a source with 
monochromatic luminosity 
$L_{\nu}(\nu)\, =\, L_{\nu}(\nu_{\rm LL}) \, (\nu/\nu_{\rm LL})^{-\alpha}$, 
where $h\nu_{\rm LL}=13.6\,{\rm eV}$, at a physical distance $r$ from the DLA 
will correspond to a ``boost factor''
\begin{equation}
\label{eqn:boost_factor}
\mathcal{B}\, =\, 15.2\, \frac{L_{\nu}(\nu_{\rm LL})}{10^{30} \, {\rm erg\,s}^{-1}\,{\rm Hz}^{-1}}\,
\frac{0.7}{\alpha} \left (\frac{r}{1\,\,{\rm Mpc}} \right)^{-2}
\end{equation}
where, for self-shielded clouds, $\mathcal{B}$ is empirically related 
to the increase in the observed \Lya\ surface brightness 
relative to that induced by the metagalactic UV background, 
SB(\Lya)$/\mu=(0.74+0.50\,\mathcal{B}^{0.89})$. 
Assuming that the observed \Lya\ emission 
is uniform across the area of sky covered 
by our observations ($0.86\times3\,{\rm arcsec}$), 
the corresponding surface brightness is 
${\rm SB(Ly}\alpha)=9.7\times10^{-18}\,{\rm erg\,s}^{-1}\,{\rm cm}^{-2}\,{\rm arcsec}^{-2}$, 
therefore requiring a boost factor $\mathcal{B} \simeq 50$.

Consider now a typical QSO at $z=1.5-2$, with spectral index 
$\alpha=1.76$ and 
$\lambda\,L_{\lambda}(\lambda)=1.5\times10^{46}\,{\rm erg\,s}^{-1}$ at $1100$\,\AA\ \citep{Tel02}. 
One can extrapolate this typical luminosity to the Lyman limit using the relation
\begin{equation}
L_{\nu}(\nu_{\rm LL})=\frac{\lambda_{\rm LL}^{2}}{c}L_{\lambda}(\lambda)
\left (\frac{\lambda}{\lambda_{\rm LL}}\right)^{2-\alpha}\approx 4\times10^{30}\,
{\rm erg\,s}^{-1}\,{\rm Hz}^{-1}.
\end{equation}
Substituting $\mathcal{B}$, $L_{\nu}(\nu_{\rm LL})$ and $\alpha$ into eq.~(\ref{eqn:boost_factor}) 
yields a physical distance $r=700\,{\rm kpc}$.
Thus, if the \Lya\ emission we see  in the spectrum of UM\,673B were
produced by a nearby source of UV photons, such a source would need to
be located within $r\sim 700\,{\rm kpc}$ 
from UM\,673,
or $\simlt 1.4\,{\rm arcmin}$ in
projection on the sky.
 As we have not identified any such source in our deep galaxy survey
of this area of sky, and none have been reported by others, we consider it
unlikely that we are observing fluorescent \Lya\ emission.

Other possibilities have been put forward \citep[e.g.][]{Dij06}, 
but all appear less likely than
the most straightforward explanation that the  \Lya\ emission we see is
produced by recombination in H\,\textsc{ii} regions ionized by 
early-type stars in a galaxy presumably associated with the DLA.
Adopting the \citet{Ken98} relationship between 
star formation rate (SFR) and \Ha\ luminosity, and assuming
the ratio Ly$\alpha$/H$\alpha \simeq  8.7$ appropriate for
case B recombination, yields an equation of the form,
\begin{equation}
{\rm SFR}\,\,(M_{\odot}\,{\rm yr}^{-1})=9.1\times10^{-43}\,L({\rm Ly\alpha})
\times \frac{1}{1.8} \,({\rm erg\,s}^{-1}).
\label{eq:sfr_lya}
\end{equation}
where the correction factor of 1/1.8 accounts for the flattening of the 
stellar initial mass function for masses below $1 {\rm M}_\odot$
\citep{Cha03} compared to the single power law
of the Salpeter IMF assumed by \citet{Ken98}.

Thus, our inferred line luminosity, 
$L({\rm Ly\alpha}) = 4.3 \times 10^{41}\,{\rm erg\,s}^{-1}$
implies a star formation rate ${\rm SFR} \simeq 0.2$\,M$_\odot$\,yr$^{-1}$.
In reality this is likely to be a lower limit, given the ease with which
\Lya\ photons are destroyed through resonant scattering in a dusty medium.
In addition, our HIRES slit may  have captured only a fraction
of the \Lya\ emission, if it is spatially more extended than the B image
of UM\,673 (a possibility which we cannot readily assess with our
spectroscopic observations).
However, we note that such low levels of star formation are not 
unusual for DLA host galaxies at $z \simlt 1$ \citep{Per10}.

\subsection{Lensed \Lya\ emission?}

We next turn to the issue of where the Ly$\alpha$ emitting 
region is located and whether it too may be lensed by the
foreground galaxy at $z=0.493$.
Since the angular diameter distances from the
lens to UM\,673 (1130 Mpc) 
and from the lens to the absorption system (1036 Mpc) 
differ by only $\sim 10\%$, we would expect the lensing
geometry to be largely unchanged.
Thus, presumably the Ly$\alpha$ photons are also  
lensed by the foreground galaxy into 
a sister image near to, but offset from, the A sightline.
The fact that we do not see any such emission in the
core of the damped Ly$\alpha$ absorption line in UM\,673A
suggests that the HIRES slit was not well placed to capture
the A counterpart of the Ly$\alpha$ emission.
A more rigorous approach involves detailed modelling of the mass
distribution of the foreground lensing galaxy, which
unfortunately is not well constrained \citep{Leh00}.

We also note that the observed Ly$\alpha$
flux may be slightly magnified, or demagnified,
by the foreground galaxy, implying
that the intrinsic Ly$\alpha$ luminosity is different from that observed.
However, a firm determination of the magnification factor is made difficult by
the uncertain mass distribution of the lensing galaxy. 
Apart from improved modelling of the lens, progress in 
the interpretation of the Ly$\alpha$ emission would be greatly facilitated
by follow-up near-infrared integral field spectroscopy aimed
at detecting the redshifted H$\alpha$ emission line at $z = 1.62650$.
Such observations would map out the full extent of the emission
region, without the limitations imposed by single-slit spectroscopy.


\begin{figure*}
  \centering
  \includegraphics[angle=0,width=155mm]{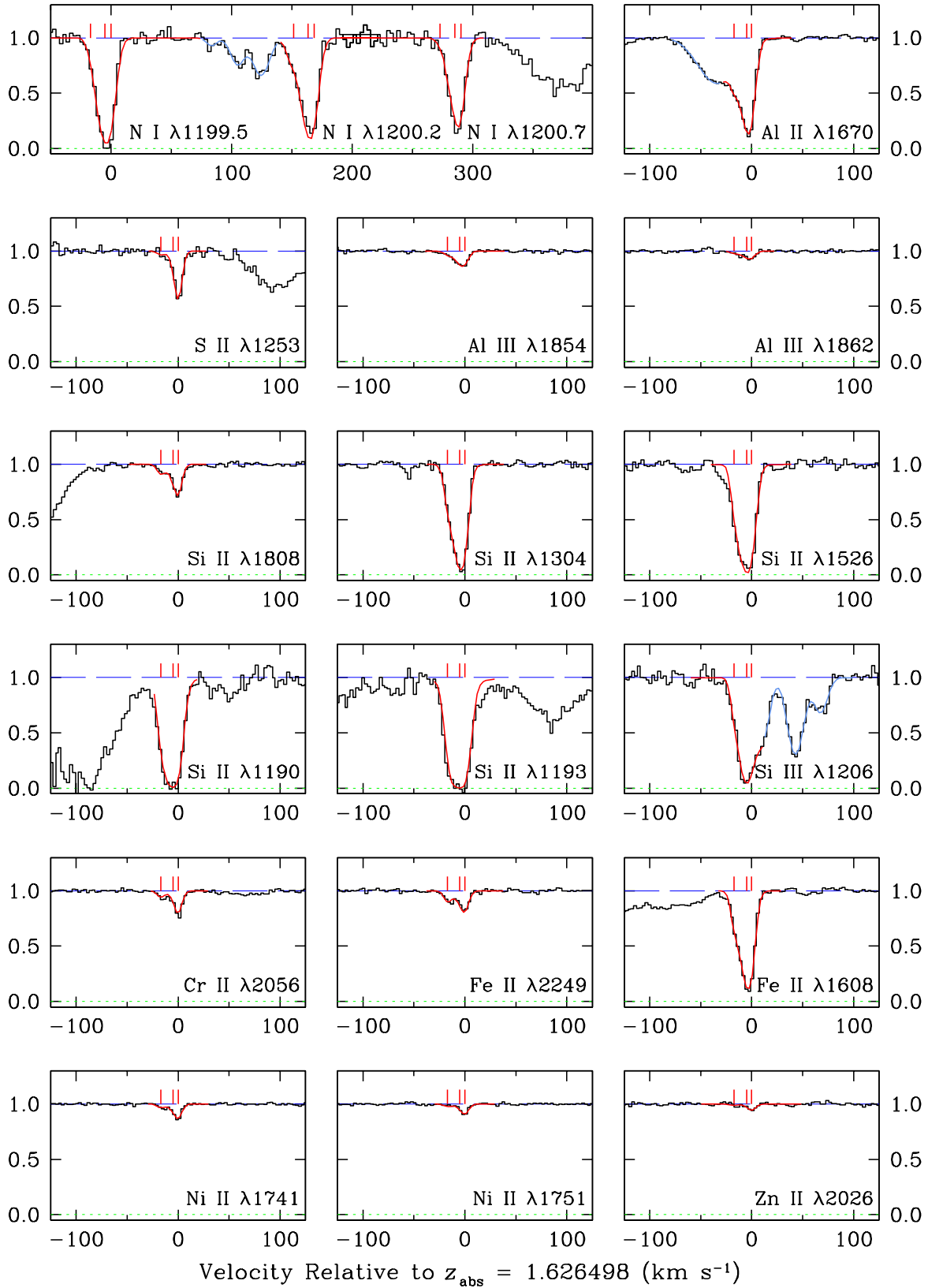}
  \caption{
  A selection of metal absorption lines associated with the DLA in UM\,673A, including
  transitions from neutral, singly and doubly ionized species. The data are shown as
  black histograms, while the red continuous lines are profile fits computed with \textsc{vpfit}
  (see text). A light blue continuous line is used to indicate nearby absorption \textit{not}
  associated with the DLA, such as the blue wing of Al\,\textsc{ii}~$\lambda 1670$
  (top right-hand panel), which has been included for completeness in the profile fitting
  procedure. The $y$-axis is residual intensity, and the velocities on the
  $x$-axis are relative to the redshift of the absorption component with the
  highest optical depth (component number 3 in Table~\ref{tab:UM673A_lowion_cloudmodel}).
  Vertical tick marks indicate the velocities of the three absorption
  components producing the absorption lines, with the parameters listed in
  Table~\ref{tab:UM673A_lowion_cloudmodel}. }
   \label{fig:UM673A_metals}
\end{figure*}

\section{Chemical Composition of the DLA in front of UM\,673A}
\label{sec:metals}

As is normally the case, a multitude
of metal absorption lines are associated with the DLA towards  UM\,673A.
With the wide wavelength coverage, high resolution and S/N ratio 
of our HIRES data, we detect 37 atomic transitions from elements
from C to Zn in a variety of ionization stages, from neutrals to triply
ionized species, as detailed in Table~\ref{tab:UM673A_lines}.
Although not all of these lines are available for abundance analysis
(some being blended or saturated), we nevertheless have at our disposal
a great deal of information on the detailed chemical composition
of this DLA. In this section, we analyse these data 
which may throw light on the chemical enrichment history of the galaxy
giving rise to the DLA and provide clues to stellar nucleosynthesis at 
low metallicities.

\begin{table}
    \caption{\textsc{Metal lines detected at the redshift of the DLA in UM\,673A}}
    \begin{tabular}{lllll}
    \hline
   \multicolumn{1}{l}{Ion}
& \multicolumn{1}{l}{Wavelength$^{\rm a}$}
& \multicolumn{1}{l}{$f^{\rm a}$}
& \multicolumn{1}{l}{$W_0^{\rm b}$}
& \multicolumn{1}{l}{$\delta W_0^{\rm b}$}\\
    \multicolumn{1}{l}{}
& \multicolumn{1}{l}{~~~~~(\AA)}
&  \multicolumn{1}{l}{}
& \multicolumn{1}{l}{(m\AA)}
& \multicolumn{1}{l}{(m\AA)} \\
    \hline
\CII      & 1334.5323      & 0.1278     & 140  & 2       \\
\CIV     & 1548.2041      & 0.1899	& \dots $^{\rm c}$        & \dots $^{\rm c}$    \\
\CIV     & 1550.7812      & 0.09475 	& 73   &  2 \\
\NI       & 1199.5496      & 0.132      & 71      & 2    \\
\NI       & 1200.2233      & 0.0869      & 67       & 2    \\
\NI       & 1200.7098      & 0.0432      & 55       & 2    \\
\OI       & 1302.1685      & 0.048       & 129    & 2    \\
\AlII     & 1670.7886      & 1.740      & \dots $^{\rm c}$        & \dots $^{\rm c}$       \\
\AlIII    & 1854.71829    & 0.559      & 14.8     & 0.9  \\
\AlIII    & 1862.79113    & 0.278      & 7.6       & 0.8 \\
\SiII      & 1190.4158      & 0.292      & 89        & 2    \\
\SiII      & 1193.2897      & 0.582      & 120      & 2  \\
\SiII      & 1260.4221      & 1.18        & 142      & 2  \\
\SiII      & 1304.3702      & 0.0863      & 83.1     & 0.9 \\
\SiII      & 1526.7070      & 0.133      & 118      & 1  \\
\SiII      & 1808.01288     & 0.00208    & 24.6     & 0.8  \\
\SiIII     & 1206.500        & 1.63      & \dots $^{\rm c}$        & \dots $^{\rm c}$               \\ 
\SiIV     & 1393.76018    & 0.513        & \dots $^{\rm c}$        & \dots $^{\rm c}$               \\
\SiIV     & 1402.77291    & 0.254        & \dots $^{\rm c}$        & \dots $^{\rm c}$               \\
\SII       & 1250.578        & 0.00543    & \dots $^{\rm c}$        & \dots $^{\rm c}$               \\
\SII       & 1253.805        & 0.0109      & 21     & 1  \\
\SII       & 1259.518        & 0.0166      & \dots $^{\rm c}$        & \dots $^{\rm c}$               \\
\CrII     & 2056.25693     &  0.103     & 21.6     & 0.9 \\
\CrII     & 2062.23610     & 0.0759      & 14.0     & 0.8 \\
\CrII     & 2066.16403     & 0.0512      & 10.3     & 0.9 \\
\FeII      & 1260.533        & 0.0240      & \dots $^{\rm c}$        & \dots $^{\rm c}$               \\
\FeII      & 1608.4509      & 0.0577      & 94.3     & 0.8 \\
\FeII      & 1611.20034    & 0.00138     & \dots $^{\rm c}$        & \dots $^{\rm c}$               \\
\FeII      & 2249.8768      & 0.001821   & 24      & 1 \\
\FeII      & 2260.7805      & 0.00244    & 32     & 1 \\
\NiII      & 1317.217       & 0.057       & \dots $^{\rm c}$        & \dots $^{\rm c}$               \\
\NiII      & 1370.132       & 0.056         & 8.4      & 0.9 \\
\NiII      & 1454.842       & 0.0323        & \dots $^{\rm c}$        & \dots $^{\rm c}$               \\
\NiII      & 1502.148       & 0.0133       & \dots $^{\rm c}$        & \dots $^{\rm c}$               \\
\NiII      & 1741.5531      & 0.0427      & 10.3    &  0.5 \\
\NiII      & 1751.9157      & 0.0277      & 6.0     & 0.6 \\
\ZnII      & 2026.13709    & 0.501      & 4.8      & 0.7 \\
\hline
\end{tabular}
\begin{flushleft}
$^{\rm a}$ Laboratory wavelengths and $f$-values from \citet{Mor03}\\
~~\,with updates by \citet{Jen06}.\\
$^{\rm b}$ Rest frame equivalent width and error.\\
$^{\rm c}$ Blended line.\\
\end{flushleft}
\label{tab:UM673A_lines}
\end{table}

\subsection{Profile Fitting}

Fig.~\ref{fig:UM673A_metals} shows a selection of metal absorption
lines associated with the DLA. The absorption is evidently confined
to a narrow velocity range, with even the strongest lines only extending
over ${\rm FWHM} \simeq 25$--30\,km~s$^{-1}$.
In order to deduce values for the column density $N$ (cm$^{-2}$)
and velocity dispersion parameter $b$ (km~s$^{-1}$), 
we employed the absorption line profile fitting 
software \textsc{vpfit}, 
which uses a $\chi^{2}$ minimization technique 
to fit multiple Voigt profiles simultaneously to several atomic transitions
and returns the best fitting values of $N$ and $b$ together with
the associated errors
\citep[see, for example,][]{Rix07}.
The theoretical line profiles generated by \textsc{vpfit}
are superimposed on the data in Fig.~\ref{fig:UM673A_metals}.
We now consider in turn gas of low, intermediate, and high ionization.


\begin{figure*}
  \centering
  \includegraphics[angle=0,width=155mm]{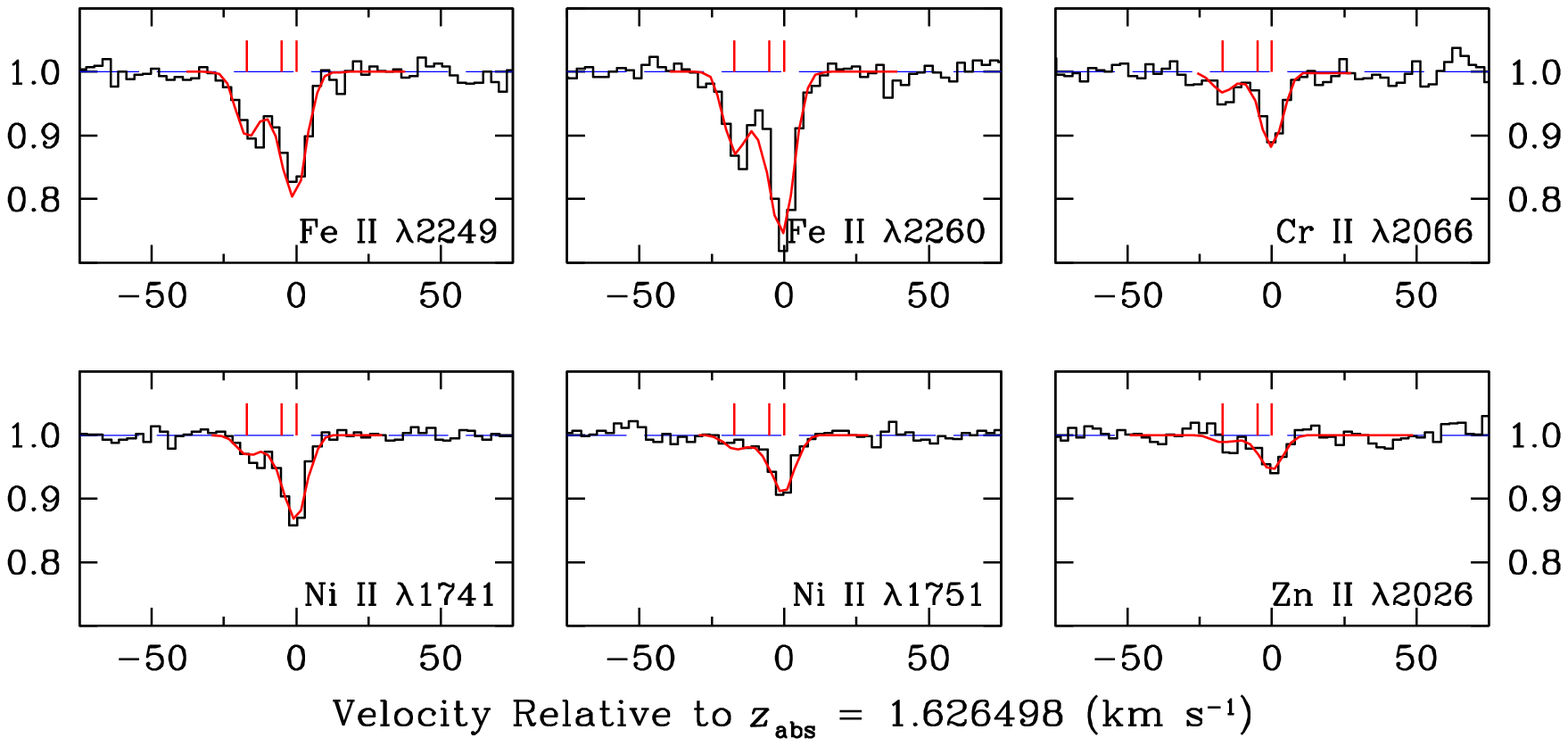}
  \caption{
  A selection of weak metal absorption lines used for
  abundance determinations in the DLA in UM\,673A,
  reproduced on an expanded scale.
  The data are shown as
  black histograms, while the red continuous lines are profile fits computed with \textsc{vpfit}
  (see text). The $y$-axis is residual intensity, and the velocities on the
  $x$-axis are relative to the redshift of the absorption component with the
  highest optical depth (component number 3 in Table~\ref{tab:UM673A_lowion_cloudmodel}).
  Vertical tick marks indicate the velocities of the three absorption
  components listed in
  Table~\ref{tab:UM673A_lowion_cloudmodel}. }
   \label{fig:weak_lines}
\end{figure*}

\subsubsection{Low ion transitions}
\label{sec:DLA_lit}

The \SiII, \CrII, \FeII\ and \NiII\ lines can all be fitted with a
minimum of three absorption components, with the parameters 
listed in Table~\ref{tab:UM673A_lowion_cloudmodel}. 
The highest optical depth is measured at $z_{\rm abs} = 1.626498$
(component number 3 in Table~\ref{tab:UM673A_lowion_cloudmodel}, or C3 for short),
which we therefore use as the zero point of the relative velocity
scale for the DLA. 
A second component (component number 1, or C1 for short)
at $z_{\rm abs} = 1.626348$, or $\Delta v = -17.1$\,km~s$^{-1}$, can
be readily recognised in the profiles of the weaker lines
(e.g. Fe\,\textsc{ii}~$\lambda 2249$) in Fig.~\ref{fig:UM673A_metals}.
What is not immediately obvious from the Figure is that a third
component (C2), 
at $z_{\rm abs} = 1.626454$ and separated by only
$\Delta v = -5.0$\,km~s$^{-1}$ from C3, is required to
reproduce the asymmetric profiles and widths of the
stronger lines. 
Note the small $b$-values, of order 1\,km~s$^{-1}$,
deduced for C1 and C3. Although these components are
unresolved with the HIRES instrumental
resolution of ${\rm FWHM} = 8.3$\,km~s$^{-1}$ (Section~\ref{sec:UM673_obs}),
their narrow widths are indicated by the relative strengths and profiles
of \SiII\ lines of widely differing oscillator strengths.
Ultimately, such low velocity dispersions can only be confirmed with
higher resolution observations. However, we note that
comparably low $b$-values are not unusual
in cool clouds in the Milky Way disk and halo \citep[e.g.][]{Pet88, Bar95},
and are now beginning to be measured at high redshifts too as 
the quality of the spectroscopic data improves \citep[e.g.][]{Jor09}.

\begin{table}
\centering
    \caption{\textsc{Absorption Components of Low Ion Transitions in the DLA in line to UM673A}}
    \begin{tabular}{@{}ccccc}
    \hline
  \multicolumn{1}{c}{Comp.}
& \multicolumn{1}{c}{$z_{\rm abs}$}
& \multicolumn{1}{c}{$\Delta v^{\rm a}$}
& \multicolumn{1}{c}{$b$}
& \multicolumn{1}{c}{Fract.$^{\rm b}$}\\
  \multicolumn{1}{c}{No.}
& \multicolumn{1}{c}{}
& \multicolumn{1}{c}{(km~s$^{-1}$)}
& \multicolumn{1}{c}{(km~s$^{-1}$)}
& \multicolumn{1}{c}{}\\
    \hline
1 & $1.626348 \pm 2 \times 10^{-6}$    & $-17.1$    & $0.68 \pm 0.07$  &  0.17 \\
2 & $1.626454 \pm 2 \times 10^{-6}$    & ~$-5.0$    & $5.9 \pm 0.2$     &  0.25 \\
3 & $1.626498 \pm 8 \times 10^{-6}$    & ~~~0.0     & $1.5 \pm 0.3$     &  0.58 \\
    \hline
    \end{tabular}

\flushleft{$^{\rm a}${Velocity relative to $z_{\rm abs} = 1.626498$}}\\

$^{\rm b}${Fraction of the total column density of Si\,{\sc ii}.}
    \label{tab:UM673A_lowion_cloudmodel}

\end{table}

With the redshift $z$ and velocity dispersion parameter $b$ fixed to be
the same for all lines of neutral and singly ionized species, 
we let the column density in each component be the free parameter
to be determined by \textsc{vpfit} [with the obvious restriction
that all absorption lines arising from the same ground state
of a given ion X{\sc n} should yield the same value of $N$(X{\sc n})].
Values of $N$(X{\sc n}) are collected in Table~\ref{tab:UM673A_lowion_coldens},
where the column densities of components C2 and C3 are
grouped together, as these two components are always blended at
the resolution of our data. 
We also list in this Table the total (C1+C2+C3)
column densities of each ion, which are better determined
than those of the individual components. 

\begin{table}
\centering
    \caption{\textsc{Ion column densities of the DLA in UM673A}}
    \begin{tabular}{@{}lccc}
    \hline
  \multicolumn{1}{l}{Ion}
& \multicolumn{1}{c}{log $N$(X)/${\rm cm}^{-2}$}
& \multicolumn{1}{c}{log $N$(X)/${\rm cm}^{-2}$}
& \multicolumn{1}{c}{log $N$(X)/${\rm cm}^{-2}$}\\
  \multicolumn{1}{l}{}
& \multicolumn{1}{c}{C1$^{\rm a}$}
& \multicolumn{1}{c}{C2+C3$^{\rm a}$}
& \multicolumn{1}{c}{C1+C2+C3$^{\rm a}$}\\
    \hline
\HI    &  N/A               &  N/A               &  $20.7\pm0.1$    \\
\NI    &  12.17 $\pm$ 0.41  &  14.97 $\pm$ 0.25  &  14.97 $\pm$ 0.25 \\
\AlII  &  11.91 $\pm$ 0.16  &  12.89 $\pm$ 0.07  &  12.93 $\pm$ 0.08 \\
\AlIII &  11.05 $\pm$ 0.16  &  11.89 $\pm$ 0.03  &  11.95 $\pm$ 0.05 \\
\SiII  &  13.99 $\pm$ 0.07  &  14.67 $\pm$ 0.02  &  14.75 $\pm$ 0.03 \\
\SiIII &  12.32 $\pm$ 0.43  &  13.05 $\pm$ 0.09  &  13.12 $\pm$ 0.17 \\
\SII   &  12.90 $\pm$ 0.43  &  14.52 $\pm$ 0.09  &  14.53 $\pm$ 0.10 \\
\TiII  &  N/A               &  N/A               &  $\leq11.90^{\rm b}$     \\
\CrII  &  12.04 $\pm$ 0.10  &  12.69 $\pm$ 0.03  &  12.78 $\pm$ 0.04 \\
\FeII  &  14.05 $\pm$ 0.05  &  14.44 $\pm$ 0.02  &  14.59 $\pm$ 0.03 \\
\NiII  &  12.17 $\pm$ 0.12  &  12.92 $\pm$ 0.02  &  12.99 $\pm$ 0.04 \\
\ZnII  &  10.54 $\pm$ 0.43  &  11.37 $\pm$ 0.09  &  11.43 $\pm$ 0.15 \\
    \hline
    \end{tabular}
\begin{flushleft}
$^{\rm a}${C1/C2/C3: Component 1/2/3.}\\
$^{\rm b}${$ 3 \sigma$ upper limit.}\\
\end{flushleft}
   \label{tab:UM673A_lowion_coldens}
\end{table}


\begin{figure*}
  \centering
  \includegraphics[angle=0,width=115mm]{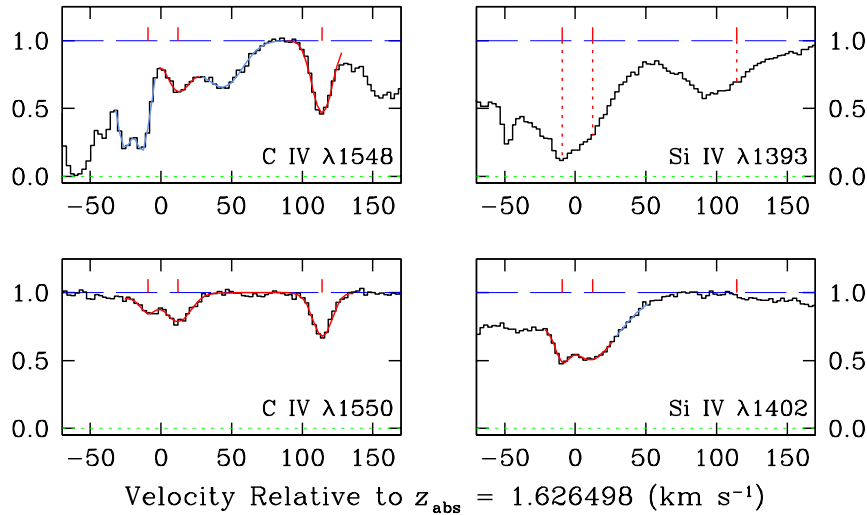}
  \caption{
   Transitions from highly ionized gas at redshifts close to that
   of the DLA in UM\,673A.
   In each plot, 
  the $y$-axis is residual intensity, and the velocities on the
  $x$-axis are relative to the redshift of the absorption component with the
  highest optical depth \textit{in neutral gas}
  (component number 3 in Table~\ref{tab:UM673A_lowion_cloudmodel}).
  The data are shown as
  black histograms, while the red continuous lines are profile fits computed with 
  the model parameters listed in Table~\ref{tab:UM673A_highion_cloudmodel}.
  A light blue continuous line is used to indicate nearby absorption \textit{not}
  associated with the DLA, but which has been included for completeness 
  in the profile fitting procedure. 
  The relative velocities of the three components of the model are indicated by red
  tick marks above the continuum level.
  All three components are blended with other absorption lines
  in \SiIV\,$\lambda1393$ (top right-hand panel), but component
  number 3, at $\Delta v = + 114$\,km~s$^{-1}$, is clearly absent
  in \SiIV\,$\lambda1402$.
   }
   \label{fig:highion}
\end{figure*}

It is important to stress in this regard that,
with the exception of \NI, our data include at least
one weak, unsaturated, transition for all of the species
used in our abundance determinations. 
Some examples are shown on an expanded scale 
in Fig.~\ref{fig:weak_lines}.
Under these circumstances,
the values of column density deduced do \emph{not} depend on the
details of the profile fitting, because the lines lie on the
linear part of the curve of growth. Although we record
\CII\,$\lambda 1334$ and \OI\,$\lambda 1302$, these 
transitions are saturated; these species are therefore
not included in Table~\ref{tab:UM673A_lowion_coldens}
and in the subsequent abundance analysis.
We do however include \TiII, whose 
$\lambda\lambda 1910.61,1910.95$ doublet is undetected;
the $3 \sigma$ upper limit $\log [N$(\TiII)/cm$^{-2}] \leq 11.90$
in Table~\ref{tab:UM673A_lowion_coldens} was deduced assuming
these lines to be as wide as \ZnII\,$\lambda 2026$, which
is the weakest feature we detect.

The column densities of the neutrals and first ions in
Table~\ref{tab:UM673A_lowion_coldens} allow us to deduce directly
the abundances of the corresponding elements.
Before doing so in Section~\ref{sec:abundances} below, we briefly
comment on the absorption from more highly ionized gas.

\begin{table}
\centering
    \caption{\textsc{Absorption Components of High Ion Transitions at Redshifts Close to That of the DLA in 
    UM\,673A}}
    \begin{tabular}{@{}ccccc}
    \hline
  \multicolumn{1}{c}{Comp.}
& \multicolumn{1}{c}{$z_{\rm abs}$}
& \multicolumn{1}{c}{$\Delta v^{\rm a}$}
& \multicolumn{1}{c}{$b$}
& \multicolumn{1}{c}{Fract.$^{\rm b}$}\\
  \multicolumn{1}{c}{No.}
& \multicolumn{1}{c}{}
& \multicolumn{1}{c}{(km~s$^{-1}$)}
& \multicolumn{1}{c}{(km~s$^{-1}$)}
& \multicolumn{1}{c}{}\\
    \hline
1 & $1.626416 \pm 3 \times 10^{-6}$    & $-9.4$         & $4.2 \pm 0.9$     &  0.51 \\
2 & $1.626605 \pm 3 \times 10^{-6}$    & $+12.2$       & $11.4 \pm 0.7$   &  0.11 \\
3 & $1.627498 \pm 2 \times 10^{-6}$    & $+114.1$     & $7.9  \pm 0.3$    &  0.38 \\
    \hline
    \end{tabular}

\flushleft{$^{\rm a}${Velocity relative to $z_{\rm abs} = 1.626498$}}\\

$^{\rm b}${Fraction of the total column density of C\,{\sc iv}.}
    \label{tab:UM673A_highion_cloudmodel}
\end{table}

\subsubsection{Intermediate ionization stages}

Our data cover two second ions, \AlIII\,$\lambda \lambda 1854, 1862$
and \SiIII\,$\lambda 1206$; all three transitions are shown in
Fig.~\ref{fig:UM673A_metals}. The weak \AlIII\ doublet lines
are well reproduced by the same `cloud model' determined
for the low ionization species (Table~\ref{tab:UM673A_lowion_cloudmodel}).
The stronger \SiIII\,$\lambda 1206$ line shows additional
redshifted absorption which presumably arises in ionized gas, since it
is absent from lines of comparable strength of ions which are dominant
in H\,{\sc i} regions (e.g. \SiII\,$\lambda 1526$---see Fig.~\ref{fig:UM673A_metals}).
Accordingly, the column densities of \AlIII\ and \SiIII\ listed
in Table~\ref{tab:UM673A_lowion_coldens} refer only to the velocity
interval appropriate to the neutral gas. These values of column densities
are useful for constraining the magnitude of putative
ionization corrections to the abundance determinations,
as discussed in Appendix~A.

\begin{table}
\centering
    \caption{\textsc{High Ion Column Densities at Redshifts Close to That of the DLA in 
    UM\,673A}}
    \begin{tabular}{@{}cccc}
    \hline
  \multicolumn{1}{c}{Ion}
& \multicolumn{1}{c}{log $N$(X)/${\rm cm}^{-2}$}
& \multicolumn{1}{c}{log $N$(X)/${\rm cm}^{-2}$}
& \multicolumn{1}{c}{log $N$(X)/${\rm cm}^{-2}$}\\
  \multicolumn{1}{c}{}
& \multicolumn{1}{c}{C1$^{\rm a}$}
& \multicolumn{1}{c}{C2$^{\rm a}$}
& \multicolumn{1}{c}{C3$^{\rm a}$}\\
    \hline
\CIV    & 12.55 $\pm$ 0.06 & 13.10 $\pm$ 0.02 & 13.22 $\pm$ 0.01 \\
\SiIV   & $12.59^{\rm b}$  & $12.45^{\rm b}$  &       $\leq 11.75^{\rm c}$        \\
    \hline
    \end{tabular}
\begin{flushleft}
$^{\rm a}${C1/C2/C3: Component 1/2/3.}\\
$^{\rm b}${Uncertain because of blending.}\\
$^{\rm c}${$3 \sigma$ upper limit.}\\
\end{flushleft}
   \label{tab:UM673A_highion_coldens}
\end{table}

\subsubsection{High ions}

We also find lines from highly ionized gas at redshifts close to, but not
the same as, that of the DLA. The extensive work by \citet{Fox07}
has shown this to be the case in many DLAs. Our HIRES spectrum
includes absorption lines from the 
\CIV\,$\lambda\lambda1548,1550$ and \SiIV\,$\lambda\lambda1393,1402$ doublets,
although out of these four lines only 
\CIV\,$\lambda 1550$ is not blended and therefore affords the
clearest view of the kinematic structure of the highly ionized gas
(see Fig.~\ref{fig:highion}).

The highly ionized gas appears to be spread over three velocity
components (Table~\ref{tab:UM673A_highion_cloudmodel});
two are close in redshift to the DLA itself 
($\Delta v = -9.4$ and +12.2\,km~s$^{-1}$ respectively
for components 1 and 2 in Table~\ref{tab:UM673A_highion_cloudmodel}),
but the third is redshifted by $\Delta v = +114$\,km~s$^{-1}$.
This third component must be of high ionization indeed,
as it is the strongest in \CIV\ and yet is absent in
\SiIV\ (see Fig.~\ref{fig:highion}).
Comparison of the $b$-values in Tables~\ref{tab:UM673A_highion_cloudmodel}
and \ref{tab:UM673A_lowion_cloudmodel} shows that
the high ions have larger velocity dispersions than
the neutral gas. 
Table~\ref{tab:UM673A_highion_coldens} lists the column densities
of \CIV\ and \SiIV.

Comparing Fig.~\ref{fig:highion} with the lower three panels of
Fig.~\ref{fig:um673blya}, it can be readily appreciated that,
\emph{in stark contrast with the low ion absorption lines}, the
\CIV\   and \SiIV\  absorption lines show little variation
between UM\,673A and B.  
Applying the same \textsc{vpfit} analysis as above to the \CIV\ lines
in UM\,673B, returns values of redshift
for the three absorption components which differ by less
than 5\,km~s$^{-1}$ from those listed in 
Table~\ref{tab:UM673A_highion_cloudmodel},
and values of column density $N$(\CIV) which differ
by less than a factor of 3 from those listed in 
Table~\ref{tab:UM673A_highion_coldens}.
The finding that highly ionized gas has a much larger
coherence scale than that of DLAs is not surprising,
and in line with the results of
earlier work on other QSO pairs 
\citep[e.g.][and references therein]{Rau01,Ell04}
and more recently on galaxy-galaxy pairs \citep{Ste10}.

\subsection{Element Abundances}
\label{sec:abundances}

\begin{figure}
  \centering
  \includegraphics[angle=0,width=80mm]{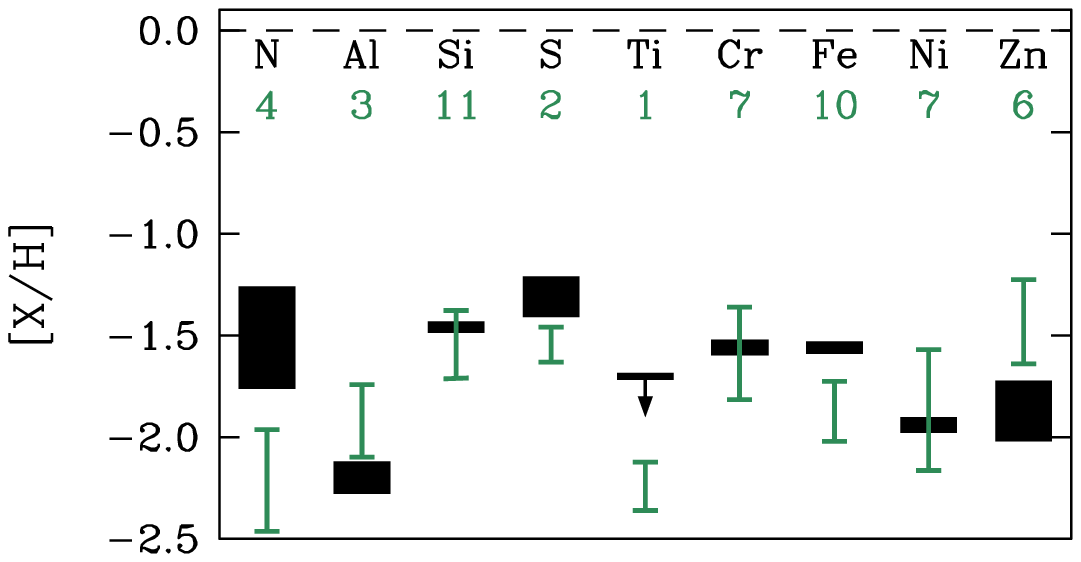}
  \includegraphics[angle=0,width=80mm]{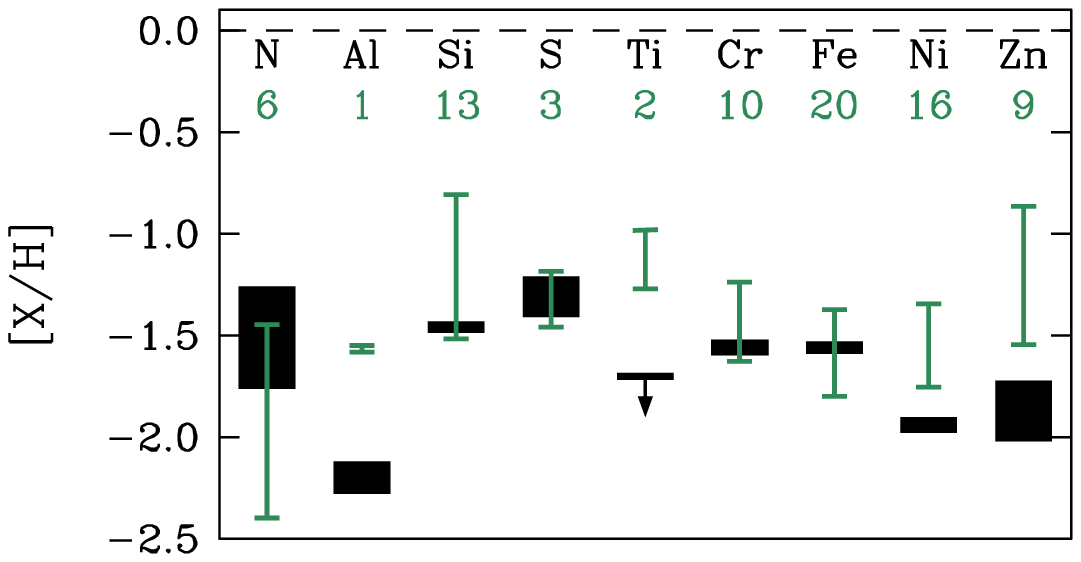}
  \includegraphics[angle=0,width=80mm]{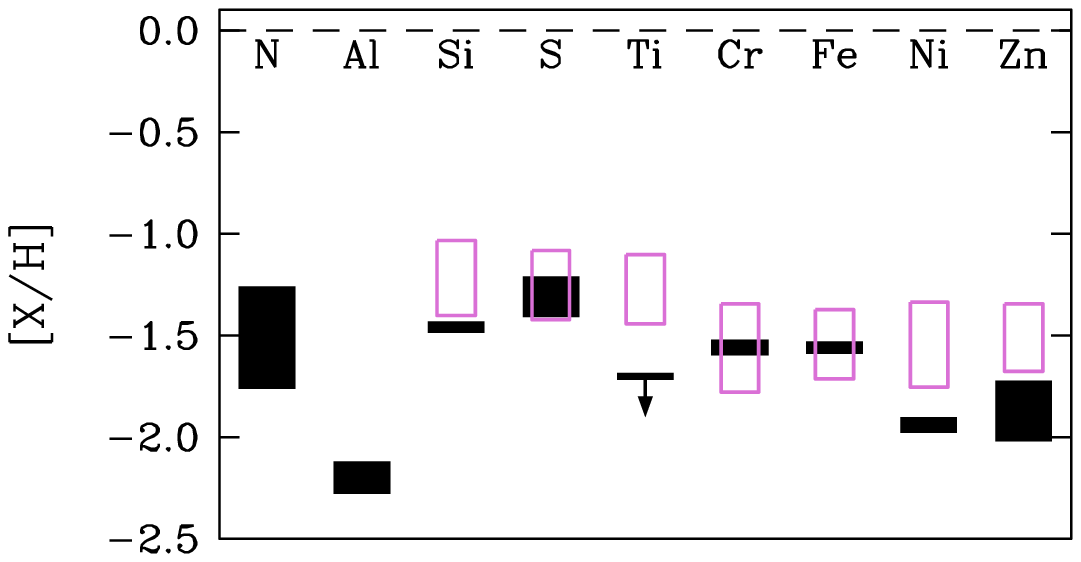}
  \medskip
  \caption{
  All three panels show the abundances of the nine elements indicated at the top in the 
  $z_{\rm abs} = 1.62650$ DLA in front of UM\,673A (black boxes);
  the size of each box corresponds to the uncertainty in each determination
  (except for Ti, for which we only have an upper limit).
   The green vertical lines 
    in the \emph{top} and \emph{middle} panels show the 
    dispersions  of the abundances of each element in samples
    of DLAs that have, respectively,  
    [Si/H] and [Fe/H] within a factor of 2 of the DLA in UM\,673A. 
    The numbers below the element tags indicate the number of DLA measurements 
    available for each element (see text for further details). 
    In the \emph{bottom} panel the DLA chemical composition is compared with that of 
    an `average' population of Galactic metal-poor stars (open boxes)
    that have [Fe/H] within a factor of 2 of the DLA.
   }
   \label{fig:metals}
\end{figure}

Abundance measurements for the DLA in UM\,673A
are collected in Table~\ref{tab:UM673A_el_abunds}
and shown graphically in Fig.~\ref{fig:metals}.
These values were deduced directly by dividing the column densities
of ions which are dominant in \HI\ regions by $N$(\HI)
(see Table~\ref{tab:UM673A_lowion_coldens}),
with the implicit assumption that corrections for ionized gas and unseen
ion stages are negligible so that, for example,
$N({\rm \SiII})/N(\HI) \equiv {\rm Si/H}$.
The validity of this assumption is examined in Appendix A,
with the conclusion that it is likely to be accurate to
within $\sim 0.05$\,dex for most elements considered here,
except Al and S, whose true abundances may be higher than their
entries in Table~\ref{tab:UM673A_el_abunds} by 0.1\,dex and 0.2\,dex respectively.
We used as reference the compilation of solar abundances by
\citet{AspGreSau05}; the very recent revision 
of the solar scale by \citet{Asp09}
differs for the elements listed in Table~\ref{tab:UM673A_el_abunds}
by at most 0.05\,dex (for N) and more generally by only 0.02--0.03\,dex.

The first conclusion to be drawn from Fig.~\ref{fig:metals}
is that the $z_{\rm} = 1.62650$ DLA in line to UM\,673A is metal-poor,
with an overall metallicity $Z \simeq 1/30 Z_\odot$, or $-1.5$ on a log
scale. Such low metallicity is not unexpected, given the narrow widths
of the absorption lines \citep[which are actually significantly narrower
than expected on the basis of the relationship proposed by][]{Pro08}
and the apparently low level of star formation deduced in Section~\ref{sec:lya}.

\begin{table}
\centering
    \caption{\textsc{Element Abundances in the DLA at $z_{\rm abs}=1.626498$ towards UM\,673A}}
    \begin{tabular}{ccc}
    \hline
  \multicolumn{1}{c}{Element}
& \multicolumn{1}{c}{log (X/H)$_{\odot}^{\rm a}$}
&  \multicolumn{1}{c}{[X/H]$_{\rm DLA}^{\rm b}$}
\\
  \multicolumn{1}{c}{(X)}
& \multicolumn{1}{c}{}
&  \multicolumn{1}{c}{}
\\
    \hline
N	& $-4.22$	& $-1.51\pm0.25$	\\
Al	& $-5.57$	& $-2.20\pm0.08$	\\
Si	& $-4.49$	& $-1.46\pm0.03$	\\
S	& $-4.86$	& $-1.31\pm0.10$	\\
Ti	& $-7.10$	& $<-1.70$			\\
Cr	& $-6.36$	& $-1.56\pm0.04$	\\
Fe	& $-4.55$	& $-1.56\pm0.03$	\\
Ni	& $-5.77$	& $-1.94\pm0.04$	\\
Zn	& $-7.40$	& $-1.87\pm0.15$	\\
\hline
\end{tabular}
\begin{flushleft}
$^{\rm a}${\citet{AspGreSau05}}\\
$^{\rm b}${$[{\rm X}/{\rm H}]_{\rm DLA}  \equiv \log({\rm X}/{\rm H})_{\rm DLA} -\log({\rm X}/{\rm H})_{\odot}$}\\
\end{flushleft}
\label{tab:UM673A_el_abunds}
\end{table}


When we look more closely at the relative abundances of the
elements measured, however, we arrive at the interesting
conclusion that this DLA exhibits some notable differences 
from the `average' populations of DLAs and 
Galactic halo stars of similar metallicity, as we now point out.
We consider two `average' sets of DLAs, each assembled from 
the HIRES DLA database compiled by \citet{Pro07}. 
In each case we selected from the database all DLAs whose abundance
is within a factor of 2 of a reference element, using in turn
Si and Fe as the reference. 
Thus, the green vertical lines in the top panel
of Fig.~\ref{fig:metals} show the dispersion in the
abundances of the elements considered here in all DLAs  with 
$-1.76\le{\rm [Si/H]}\le-1.16$, i.e. within a factor of $\sim 2$
of the value ${\rm [Si/H]} = -1.46 \pm 0.03$ we measure
towards UM\,673A.
The middle panel shows the same data for all
DLAs with $-1.86\le{\rm [Fe/H]}\le-1.26$. 
The numbers below the element tags in Fig.~\ref{fig:metals}
indicate the number of DLAs
included in the average population 
for the corresponding element. 
If there exists only one measurement, 
the size of the error bar represents the uncertainty 
in that single measurement. 
If just two or three measurements exist, 
we instead plot the standard error in the mean. 
Otherwise, we use a robust determination of the dispersion in the average population
(see Section~\ref{sec:sizes}).

Similarly, in the bottom panel of Fig.~\ref{fig:metals}, 
we overlay (empty boxes) the element abundances of the
average population of Galactic metal poor stars drawn 
from the samples of \citet{Gra03} and \citet{Nis07}, using Fe as the reference
element. For all available measurements, 
the height of the box represents the dispersion 
in the average population.

Considering first the top panel of Fig.~\ref{fig:metals},
it appears that the DLA in UM\,673A is Fe-rich and 
Zn-poor relative to other DLAs with similar Si abundance.
There may be offsets in N and S too, but their statistics
are poorer. The best DLA statistics are those for
the middle panel, where we see that, relative to other
DLAs with similar Fe abundance, the absorber in UM\,673A
is deficient in Ti, Ni, and Zn. The same conclusion is
reached by comparing with the stellar abundances in
the bottom panel.

It seems unlikely that these apparently anomalous abundances
are due to dust depletion, given that:
(a) Ti, Ni, and Zn are depleted
to very different degrees in Galactic dust \citep{SavSem96},
and (b) depletions are in any case expected to introduce
very minor corrections when the overall metallicity is as low
as 1/30 of solar \citep[e.g.][]{Ake05}.

\begin{figure}
  \centering
  \hspace{-0.2cm}\includegraphics[angle=0,width=80mm]{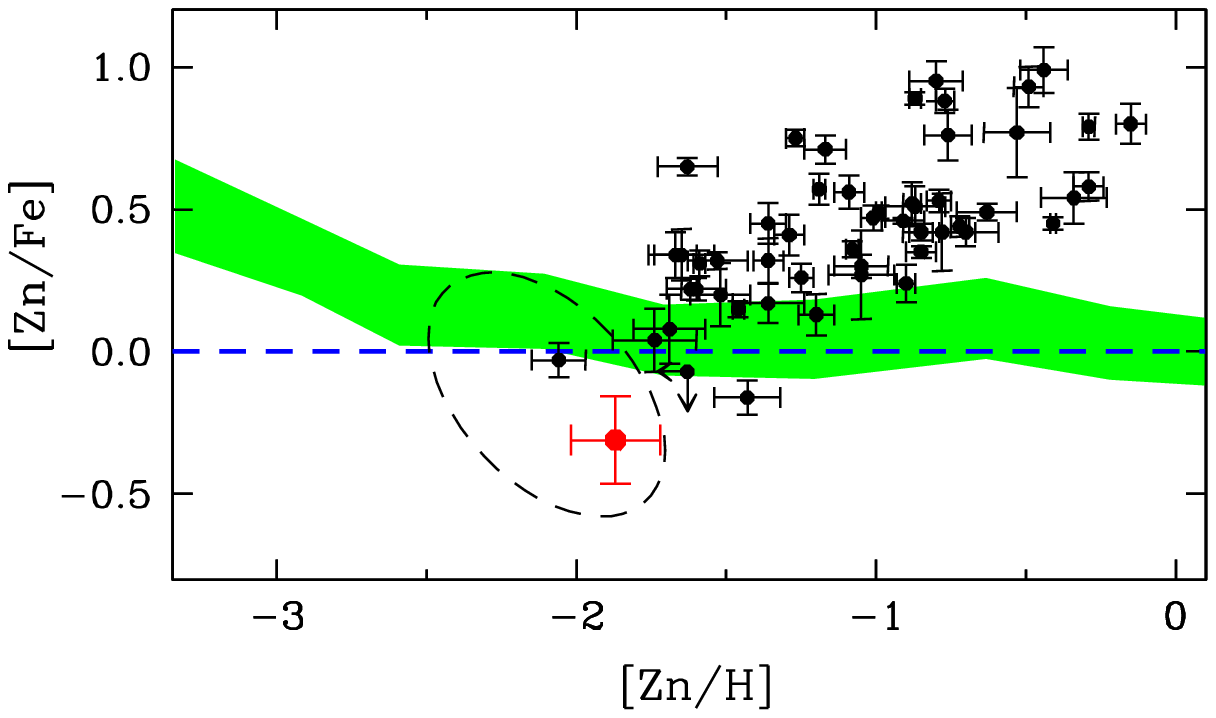}
  \caption{
  Zn and Fe abundances in DLAs from the compilations by \citet{Pro07},
  \citet{Des07}, and \citet{Not08} are
  shown with black circles, while the red dot is for the DLA in UM\,673A.
  Note that the fall off in [Zn/Fe] with decreasing [Zn/H] in DLAs
  is not thought to be intrinsic, but rather to reflect a decreasing fraction of Fe
  in dust grains at lower overall metallicities, as measured by the 
  normally undepleted Zn.
  The green shaded region shows the behaviour of [Zn/Fe] in Galactic stars
  \citep[][and references therein]{Sai09}, 
  with the width of the band corresponding to the $\pm 1\sigma$ 
  dispersion of values at a given metallicity. 
  The dashed ellipse indicates the approximate locus of 
  stars in the dwarf spheroidal galaxies Draco,
  Ursa Minor, and Sextans from the work by 
  \citet{SheCotSar01} and \citet{CohHua09}.
  }
   \label{fig:zn}
\end{figure}

The unusually low abundance of Zn
\citep[an element which is not readily incorporated into dust grains and thus normally
provides a reliable benchmark for the abundance of Fe-peak elements, e.g.][]{PetBokHun90},
can best be appreciated from Fig.~\ref{fig:zn}, where the value
of [Zn/Fe] in UM\,673A is compared with those measured in DLAs
and Galactic stars. In the Figure, 
the trend of decreasing [Zn/Fe] with decreasing
[Zn/H] in DLAs is most naturally explained by the 
reduced dust depletion of Fe at lower metallicities,
until solar relative abundance of the two elements 
is recovered when ${\rm [Zn/H]} \simeq -1.5$.
In Galactic stars, Zn and Fe track each other closely
over two decades in metallicity (from solar
to $\sim 1/100$ solar), and Zn becomes progressively \emph{over}abundant
relative to Fe with decreasing metallicity when [Zn/H]\,$\simlt -2$.
Evidently, the [Zn/Fe] ratio is lower in UM\,673A
than in any other DLA in current samples; similarly,
none of the metal-poor stars in 
the compilations by \citet{Nis07} and \citet{Sai09}
exhibits an underabundance of Zn relative to Fe as pronounced
as that uncovered here.

We have searched for clues in the composition of Galactic stars
that are much more metal-poor than 1/30 of solar, in the
extreme regime where chemical anomalies due to enrichment by only a few
prior episodes of star formation 
may manifest themselves, but found none.
The works by \citet{Cay04} and \cite{Lai08} show 
that when ${\rm Fe/H} \ll -2$,
Ti is \emph{over}abundant by a factor of $\sim 2$ relative to Fe,
Cr is progressively \emph{under}abundant with decreasing [Fe/H],
Ni/Fe remains solar, and Zn exhibits the behaviour illustrated
in Fig.~\ref{fig:zn}, quite unlike the relative abundances
of these four elements relative to Fe  in the DLA towards UM\,673A.


Recently, data have become available on the abundances of these elements
in individual stars of dwarf spheroidal galaxies in the Local Group
\citep{SheCotSar01,CohHua09} and in ultra-faint dwarf companions 
of the Milky Way \citep{Fre10}. These measurements are of particular
interest, as some of these galaxies may well have experienced different 
star formation histories from the stellar population(s) of the Galactic halo,
and may therefore offer a different perspective for the interpretation 
of element ratios. 
It is thus intriguing to find that stars with [Fe/H]\,$\simlt -2$
in the Draco, Sextans, and Ursa Minor dwarf spheroidal galaxies
can exhibit sub-solar [Zn/Fe] ratios (see Fig.~\ref{fig:zn}), and 
solar or sub-solar [Ti/Fe] (in contrast with the super-solar [Ti/Fe]
of Galactic halo stars). 
However, the resemblance to the DLA in UM\,673A
does not extend to other elements: in the dwarf spheroidal stars
observed by \citet{SheCotSar01} and \citet{CohHua09} the mean 
value of [Ni/Fe] is approximately solar and [Cr/Fe] is mostly
subsolar, whereas the opposite is found in the DLA (see Fig.~\ref{fig:metals}).
Thus, while these data do offer examples of clear
departures of some element ratios from the well-established
Galactic halo pattern, a definite chemical correspondence 
between the DLA in UM\,673A and nearby dwarf spheroidal
galaxies cannot be established.

It is possible that, as more data of comparably high S/N ratio 
to those presented here are obtained for other very metal-poor
DLAs, more examples of anomalous element abundances will
be uncovered, challenging current calculations of stellar yields
at low metallicities. Sub-solar [Zn,Ni,Ti/Fe] ratios in 
very metal-poor DLAs may well be more common than has
been appreciated so far, because the absorption lines
of all three elements become vanishingly small 
when [Fe/H]\,$\simlt -1.5$ \citep[see, for example,][]{Pet08a}, 
requiring data of unusually high S/N ratio for their abundances to
be determined.

\section{Summary and Conclusions}
\label{sec:conc}

Thanks to the high efficiency of the HIRES spectrograph at ultraviolet wavelengths,
we have uncovered a damped \Lya\ system at $z_{\rm} = 1.62650$ in the spectrum
of the gravitationally lensed QSO UM\,673; 
the DLA had been overlooked until now, despite
the many observations of this bright QSO over the last quarter of a century.
From the analysis of high resolution and S/N ratio spectra of each image of the 
UM\,673A,B pair,
we draw the following conclusions.\\

\noindent ~(i) In the direction probed, 
the transverse extent of the DLA is much less than 
2.7\,kpc (this being the separation of the two sightlines at $z =  1.62650$), since
  we measure a drop by a factor of at least 400 in the column density
  of neutral hydrogen from UM\,673A
  ($\log [N$(\HI)/cm$^{-2}] = 20.7$) to UM\,673B ($\log [N$(\HI)/cm$^{-2}] \leq 18.1$).
  A comparable drop is seen in the column density of \SiII\ and presumably
  other metals in stages that are dominant  in \HI\ regions. 

\smallskip

\noindent ~~(ii) From a reassessment of data on other QSO pairs in the literature, together
  with the new results for UM\,673A,B, we find that, if the radial profile of $N$(\HI) 
  in DLAs declines exponentially, the median $e$-folding scale length 
  is $\tilde{\rm S}_{\rm DLA,e}=1.3 \pm 0.8\,{\rm kpc}$,
  smaller than had previously been realised.
  For a spherical cloud, this corresponds to a typical DLA radius $R\simeq 5 \pm 3$\,kpc.  

\smallskip

\noindent  ~~(iii) Towards UM\,673B, 
  we detect a weak and narrow \Lya\ \emph{emission} 
  line at the same redshift as the DLA in UM\,673A. 
  If the line is due to 
  recombination in \HII\ regions,
  which we consider to be the most likely interpretation,  
  its low luminosity ($L({\rm Ly\alpha}) = 4.3 \times10^{41}\,{\rm erg\,s}^{-1}$)  
  implies a modest star formation rate,  
  ${\rm SFR} \simeq 0.2$\,M$_\odot$\,yr$^{-1}$. 
  However, this is probably a lower limit
  considering: (a) the ease
  with which resonant \Lya\ photons can be destroyed or 
  scattered out of the line of sight, and (b) the limited spatial
  sampling of the narrow HIRES entrance slit. 
 
\smallskip
  
\noindent ~~(iv) In contrast with neutral gas, 
  absorption by \CIV\ exhibits only modest variations between 
  the two sightlines. Evidently, highly ionized gas extends over much larger physical
  dimensions than the DLA,  in accord with earlier
  conclusions from other QSO pairs and recent work on galaxy-galaxy pairs.

\smallskip
  
\noindent ~~(v)  The DLA in front of UM\,673A is metal poor, with overall
  metallicity $Z_{\rm DLA} \simeq 1/30 Z_\odot$, and has a very simple
  velocity structure, with three absorption components spread over a narrow
  velocity interval, $\Delta v = 17$\,km~s$^{-1}$. Two of the components
  have very small internal velocity dispersions, 
  with $b = \sqrt{2} \sigma \simeq 1$\,km~s$^{-1}$,
  where $\sigma$ is the one-dimensional rms velocity of the absorbing ions
  projected along the light of sight.

\smallskip
  
\noindent ~~(vi) We are able to determine with precision the relative abundances
of nine chemical elements, from N to Zn, thanks to the large number of metal
absorption lines recorded at high S/N ratio. There appear to be some peculiarities
in the detailed chemical make-up of the DLA, with the elements Ti, Ni, and Zn
being deficient by factors of $\sim 2$--3 compared to other DLAs and to
Galactic halo stars of similar overall metallicity. 
The [Zn/Fe] ratio is the lowest measured in 
existing samples of DLAs and halo stars.
While comparably low values of [Zn/Fe] have been measured
in some stars of nearby dwarf spheroidal galaxies, other 
element ratios differ between these stars and the DLA,
so that a direct chemical correspondence cannot be established.
An interpretation of these peculiar element ratios
in terms of the previous history of 
chemical enrichment of the gas is still lacking.

Taken together, the small size, 
quiescent kinematics, and near-pristine chemical 
composition of the DLA 
in front of UM\,673A 
would suggest an origin in a low-mass galaxy.
However, the detection of nearby \lya\ emission
adds a new `twist'
to this picture. Although not intersected by our line
of sight, there must presumably be a significant mass
of cold gas within $\sim 3$\, kpc of the DLA to fuel
the star formation rate we deduce from the \lya\ luminosity.
If this is the case, then the properties we measure
may refer to an interstellar cloud---or complex
of clouds---within a larger galaxy, rather
than being representative of the whole galaxy.
This caveat may also apply to 
other DLAs whose transverse dimensions have been
probed with QSO pairs.
However, the fact that
in \emph{none} of the cases studied so far
has common DLA absorption been found over scales
of a few kpc (see Table~\ref{tab:dla_sizes})
points to one (or both) of two possibilities:
either interstellar clouds with 
$N$(\HI)\,$\geq 2 \times 10^{20}$\,cm$^{-2}$ 
have covering fractions $f_{\rm c} \ll 1$ within the 
interstellar media of high redshift galaxies or,
if $f_{\rm c} \sim 1$, the host galaxies of most
DLAs are genuinely of small extent.
The generally low metallicities of most DLAs
independently point to an origin in low luminosity galaxies
as discussed, among others, by \citet{Fyn08}.

Looking ahead, the nature of the DLA studied here 
would undoubtedly be clarified
by integral field observations of UM\,673 at the 
wavelength of the \Ha\ emission line 
(which at $z = 1.62650$ is redshifted into the 
near-infrared $H$-band, 
at $\lambda_{\rm obs} = 1.7241\,\mu$m).
Its detection would
confirm the presence of a star-forming region, its
spatial extent, and whether or not it is lensed by the
foreground galaxy.
As a closing remark, we also point out that, with its narrow
velocity width and low metallicity, the DLA in UM\,673A
is a prime candidate for a rare measurement of the primordial D/H ratio
\citep{Pet08b}. Such data, however, can only be obtained
with the \textit{Hubble Space Telescope} because the higher 
order Lyman lines in which the isotope shift could be
resolved all lie at ultraviolet wavelengths 
inaccessible from the ground.

\vspace{-0.85cm}

\section*{Acknowledgements}
We are grateful to the staff astronomers at the Keck  
Observatory for their assistance with the observations.
It is a pleasure to acknowledge advice and help with various aspects
of the work described in this paper by George Becker, 
Sebastiano Cantalupo, Bob Carswell,
Martin Haehnelt, Paul Hewett, Geraint Lewis, Eric Monier, and Sam Rix.
We thank the Hawaiian
people for the opportunity to observe from Mauna Kea;
without their hospitality, this work would not have been possible.
RC is jointly funded by the Cambridge Overseas 
Trust and the Cambridge Commonwealth/Australia Trust 
with an Allen Cambridge Australia Trust Scholarship.
CCS's research is partly supported by grants
AST-0606912 and AST-0908805 from the US National Science Foundation.
MP  would like the express his gratitude to the members of
the International Centre for Radio Astronomy Research
at the University of Western Australia for their generous 
hospitality during the progress of this work.


\begin{appendix}
\label {app:ion_corr}

\section{Ionization Corrections}

In Section~\ref{sec:abundances} we determined the chemical composition
of the DLA by assuming that the ion stage which is dominant in \HI\ regions
could be taken to represent the total column density of the corresponding element
in the DLA. This is normally a safe assumption at the high 
neutral hydrogen column densities of DLAs,
where the gas is self-shielded from ionizing radiation \citep[e.g.][]{Vla01}.
However, given the unusual abundance pattern uncovered here,
it is worthwhile reexamining the assumption that, for example, 
$N{\rm(\NI)}/N{\rm (\HI)} \equiv {\rm N/H}$, and assess 
to what degree, if any, N may be over- or under-ionized compared to
H (and the same for the other elements considered).

To this end, we ran a suite of \textsc{cloudy} photoionization models 
\citep{Fer98}, assuming that the DLA can be approximated by a slab 
of constant density gas in the range $-3<\log[n({\rm H})/{\rm cm}^{-3}]<3$. 
In our simulations, we included the  \citet{HarMad01} metagalactic
ionizing background, as well as the cosmic microwave background,
both at the redshift of the DLA. 
We adopted the solar abundance scale of \citet{AspGreSau05} 
and globally scaled the metals to $Z_{\rm DLA} = 1/30 Z_{\odot}$. 
No relative element depletions were employed, nor were grains added. 
The simulations were stopped when the column density of the DLA was reached. 
We are then able to calculate the ionization correction, IC(X),
for  element X in ionization stage \textsc{n}
by the relation
\begin{equation}
\label{eq:ics}
\mathrm{IC(X)} = \log \bigg[\frac{N(\mathrm{X})}{N(\mathrm{H})}\bigg]_{\mathrm{intrinsic}} - 
\log \bigg[\frac{N(\mathrm{X}\,\textsc{n})}{N(\mathrm{H}\,\textsc{i})}\bigg]_{\mathrm{computed}}\,\,\, 
\end{equation}
which will be negative if we overestimate the abundance of an element 
by assuming that the dominant ionization stage is representative of the true abundance. 

\begin{figure}
  \centering
  {\includegraphics[angle=0,width=80mm]{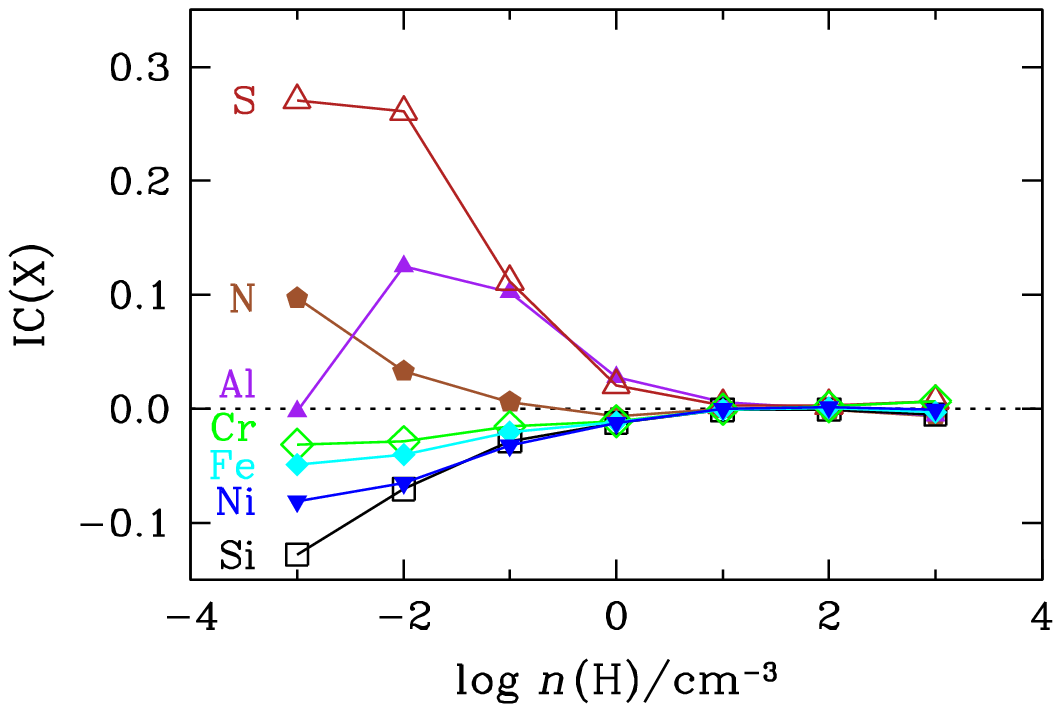}\vspace{0.5cm}}
  \includegraphics[angle=0,width=80mm]{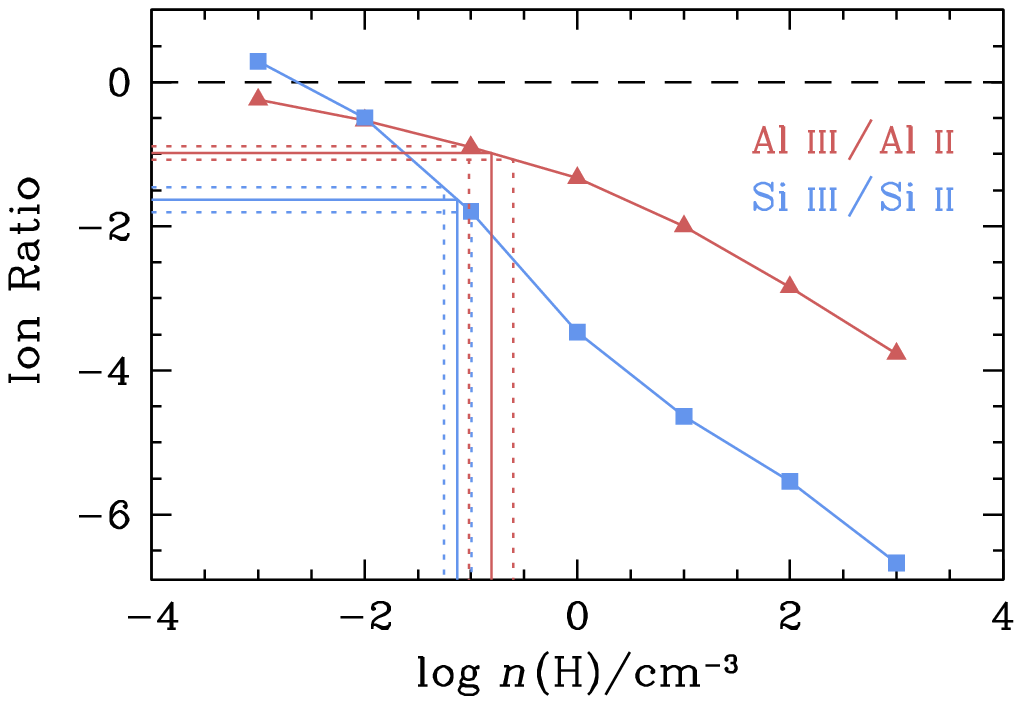}
  \caption{
   \emph{Top:} Ionization corrections, as defined by Eq.~\ref{eq:ics}, as a function of the gas density,
    $n$(H), calculated by \textsc{cloudy}
    for the elements of interest (see text for further details of the calculations).
  \emph{Bottom:} Ratios of successive ion stages of Si and Al (symbols connected by solid lines),
    as a function of $n$(H). The
    values of these ratios measured  in the DLA are indicated with straight lines; 
    the dotted lines show the corresponding  uncertainties.
   }
   \label{fig:ics}
\end{figure}

The results of this exercise are shown in Fig.~\ref{fig:ics}, where it can
be seen (top panel) that the ionization corrections are less than 0.1\,dex for most
elements considered, for gas densities in excess of $\log [n{\rm (H)/cm}^{-3}] = -2 $.
We can constrain the density by considering the relative
column densities of successive ion stages, in our case using the
\AlIII/\AlII\ and \SiIII/\SiII\ pairs. 
As can be seen from the bottom panel in Fig.~\ref{fig:ics},
both ratios give consistent answers, indicating a density
$\log [n{\rm (H)/cm}^{-3}] = -1.1 \pm 0.1 $ (from the 
better determined \SiIII/\SiII\ ratio).
At these densities, the ionization corrections for the
elements in question are smaller than the uncertainties
in the corresponding ion's column densities (shown by the
height of the black boxes in Fig.~\ref{fig:metals}), and can
thus be safely neglected.
The same conclusion is reached if the radiation field
responsible for ionizing the gas has a purely stellar
origin, rather than the mix of Active Galactic Nuclei
and star-forming galaxies
that is the source of the metagalactic background considered
by  \citet{HarMad01}.

We can estimate the line-of-sight distance
through the DLA, ${\rm D}_{\rm los}$, from our derived volume density 
under the assumption of constant density
[i.e. ${\rm D}_{\rm los} = N({\rm H}\,\textsc{i})/n({\rm H})$].
We find ${\rm D}_{\rm los}=2.1_{-0.7}^{+1.0} \, {\rm kpc}$, 
consistent with the absence of the DLA in UM\,673B and in agreement
with the characteristic sizes deduced from our analysis in Section~\ref{sec:sizes}.

\end{appendix}


\label{lastpage}

\end{document}